\documentstyle[11pt,aaspp4,psfig]{article}
\def\gs{\mathrel{\raise0.35ex\hbox{$\scriptstyle >$}\kern-0.6em % Greater/squiggles
\lower0.40ex\hbox{{$\scriptstyle \sim$}}}}
\def\ls{\mathrel{\raise0.35ex\hbox{$\scriptstyle <$}\kern-0.6em % Less than/squiggles
\lower0.40ex\hbox{{$\scriptstyle \sim$}}}}

\def\arcsper{\ifmmode \rlap.{''}\else $\rlap{.}''$\fi}
\def\arcmper{\ifmmode \rlap.{'}\else $\rlap{.}'$\fi}

\def\gs{\mathrel{\raise0.3ex\hbox{$\scriptstyle >$}\kern-0.55em 
 \lower0.60ex\hbox{{$\scriptstyle \sim$}}}}
\def\ls{\mathrel{\raise0.3ex\hbox{$\scriptstyle <$}\kern-0.55em 
 \lower0.60ex\hbox{{$\scriptstyle \sim$}}}}
\begin{document}
%\small

\title{Morphological Studies of the Galaxy Populations in Distant ``Butcher--Oemler'' Clusters with $HST$. II. AC~103, AC~118, and AC~114 at 
$z=0.31$.\footnote{Based on observations obtained with the
NASA/ESA Hubble Space Telescope which is operated by STSCI for the Association  
of Universities for Research in Astronomy, Inc., under NASA contract  
NAS5-26555.}}
\author{Warrick J.\ Couch}
\affil{School of Physics, University of New South Wales, Sydney 2052, Australia}
\author{Amy. J.\ Barger\altaffilmark{2}}
\affil{Institute of Astronomy, Madingley Rd, Cambridge CB3 OHA, UK}
\author{Ian Smail}
\affil{Department of Physics, University of Durham, South Rd, Durham 
DH1 3LE, UK}
\author{Richard S.\ Ellis}
\affil{Institute of Astronomy, Madingley Rd, Cambridge CB3 OHA, UK}

\and 

\author{Ray M.\ Sharples}
\affil{Department of Physics, University of Durham, South Rd, Durham 
DH1 3LE, UK}

\altaffiltext{2}{present address: Institute for Astronomy, University of 
Hawaii, 2680 Woodlawn Dr., Honolulu, HI  96822}

\begin{abstract}
We present new results of a program to study the detailed morphologies of 
galaxies in intermediate redshift clusters and hence understand the physical 
origin of the enhanced star formation seen in these environments at  
earlier epochs. Deep, high resolution imagery has been obtained of 3 rich
clusters, AC~103, AC~118 \& AC~114 at $z=0.31$, through the $R$(F702W) filter 
of the Wide Field Camera (WFPC-2) of the {\it Hubble Space Telescope} ($HST$). 
For AC~103 and AC~118,  single 3--orbit pointings covering a central 
$\sim 0.5\times 0.5\,h^{-1}$\,Mpc area have been obtained. For AC~114, a mosaic 
of four 6--orbit pointings has provided coverage  of a central 
$\sim 1.2\times 0.7\,h^{-1}$\,Mpc area, allowing for the first 
time a study extending to the outer regions of a more distant cluster.  
Morphological classification has been conducted visually using these images,  
with Hubble types plus evidence of dynamical interactions and/or structural
abnormalities recorded for all galaxies down to $R_{702}=22.25$ in AC~103 \&  
AC~118, and $R_{702}=23.00$ in AC~114. New spectroscopy limited at $K'\leq 18$  
has also been acquired, providing within our WFPC-2 images a total sample of 
129 confirmed cluster members of which 117 have secure star formation  
classifications (eg. starburst, post--starburst, H$\delta$--strong).  

Our study reveals that the mix of Hubble types within the distant clusters is,  
overall, significantly different to that seen in the same high--density  
environments nearby: there are proportionally more spiral galaxies and fewer 
E/S0  galaxies  at these earlier epochs. Within the core 
($r\ls 200\,h^{-1}$\,kpc) regions of the 3 clusters, spirals covering the 
full Sa--Sdm/Irr range are present in numbers up to a factor of $\sim 4$ higher 
than that seen in present--day clusters, the highest fraction being observed 
in the least massive cluster within our sample, AC~103. Only in the virialised  
core of our most massive, regular cluster, AC~114, do we see morphological  
fractions approaching those of the present epoch. However, outside the core of  
this cluster the spiral fraction rises to $\sim 60$\% -- 3 times above the  
present--day level. Dynamical interactions are also widespread throughout the
populations of all three clusters with $\sim 20$\% of the members, on average, 
showing morphological evidence of this phenomenon.  

The various subsets of `active' cluster members show emergent morphological  
trends. The majority of blue galaxies either undergoing a starburst or seen 
shortly ($<0.5$\,Gyr) thereafter are systems involved in {\it major}  
mergers. These galaxies, however, are generally of modest luminosity 
($L\sim L^{*} + 1$\,mag) even in this brightened phase; in their faded state  
they appear destined to become dwarfs, too faint to be included in  
magnitude--limited spectroscopic samples such as ours. Cluster members with  
ongoing star formation typical of present-day spirals are mostly normal  
Sb--Sdm/Irr Hubble types. Galaxies identified as having completed their last  
major episode of star formation 1--2\,Gyr prior to the epoch of observation 
are conspicuous by their commonality in morphology, all being early--type 
(S0--Sb) disk systems. The red H$\delta$--strong objects, interpreted from 
their colors and spectra as being the remnants of secondary star formation in  
old dormant systems, have morphologies consistent with this picture, being a  
mixture of normal E and S0 galaxies. 

In combination, these results point to cluster galaxy evolution being driven
by not one but several different physical processes. The only conspicuous one
is galaxy--galaxy interactions and merging. However, our study reveals 
many galaxies where this process could not have been responsible for altering
the course of their star formation activity. Furthermore, the normal disk
morphology of the majority of these previously active galaxies indicate the
operation of process(es) which halt star formation but leave the basic disk
structure intact and largely unperturbed. This tends to favour mechanisms 
which affect the gas supply (eg. ram-pressure stripping, galaxy infall) 
rather than processes like ``galaxy harassment'' where we simply do not see
the population of severely distorted spirals expected if its operation is
widespread.     

\end{abstract}


\sluginfo

\section{Introduction}

One of the longest--standing and most striking pieces of evidence for
galaxy evolution is the so--called ``Butcher-Oemler'' (BO) effect
whereby the cores of rich, regular clusters at $z\gs 0.2$ are observed
to contain many more blue galaxies than in their present--day counterparts
(Butcher \& Oemler 1978, 1984; Couch \& Newell 1984; see also Rakos \& 
Schombert 1995). Yet, understanding how cluster populations can be so different 
at recent look--back times and determining the physical processes responsible 
for their subsequent transformation remain an important challenge in  
observational cosmology. This series of papers provides an important  
contribution towards the resolution of these issues through comprehensive  
analyses of magnitude-limited samples of member galaxies in $z\sim 0.3$ 
clusters for which high quality spectra, deep infrared imaging and  
post-refurbishment $HST$ imaging has been secured. The relative proximity of  
these clusters and their varied evolutionary state as judged from their masses  
and X-ray luminosities together with the uniform nature of the datasets  
constructed, complements other analyses based on samples which span a wider  
range of redshift and include a larger number of clusters, particularly the  
``MORPHS'' sample (Smail et al. 1997a, Dressler et al. 1997a; see also 
Stanford et al. 1996).   

In the first paper of this series (Couch et al. 1994; hereafter Paper I), 
an initial indication of the morphological nature of galaxies in our distant 
cluster sample was obtained using images taken with $HST$ in its aberrated
state. Deep Wide Field and Planetary Camera (WFPC-1) images were gathered 
of two rich clusters, AC~114 at $z=0.31$ and Abell~370 at $z=0.37$, for which
comprehensive ground--based photometry and spectroscopy was available 
(Couch \& Sharples 1987; MacLaren, Ellis \& Couch 1988; Arag\'{o}n-Salamanca,  
Ellis \& Sharples 1991). These data allowed us to morphologically classify, with
moderate precision, the galaxy populations in these clusters, in particular
those active or recently active in star formation and responsible for the
BO--effect. This revealed two important trends amongst the population of
blue active galaxies: (1)\,most had a disk-dominated, late-type spiral
morphology which is rarely seen in the cores of present-day clusters, and 
(2)\,almost half appeared to be involved in dynamical interactions and mergers, 
a process first identified in distant rich clusters by Lavery \& Henry (1988; see also Lavery, Pierce \& McClure 1992). Similar trends were also found by  
Dressler et al. (1994a,b) in their separate $HST$--based study of the cluster
Cl0939+37 at $z=0.39$. Also examined in Paper I was the morphology of red  
cluster members with spectroscopic and/or photometric indications of previous  
star formation activity. In contrast to the blue population, these galaxies
appeared to be isolated and undisturbed with a normal spheroidal morphology. 
This work gave a first indication that in the BO effect we were dealing
with not only a wide-spread change in the typical star formation
properties of galaxies in rich clusters, but also with a transformation
of their morphologies.

Using spectroscopic measures, Barger et al. (1996; hereafter B96) explored in 
more detail the star formation characteristics seen amongst the distant cluster 
galaxies in order to better interpret the morphologies observed for these 
objects. They showed that the {\it numbers} of galaxies found in the 
different phases of star formation activity -- starburst, post-starburst,  
spiral-like, red H$\delta$-strong (Couch \& Sharples 1987, hereafter CS; 
Paper I) -- were consistent with a model incorporating a secondary starburst 
in $\sim 30$\% of the distant cluster populations. When B96 combined this with
preliminary morphologies derived from a subset of the $HST$ data we present
here, there was a strong suggestion that mergers and interactions played an
important part in triggering this star formation cycle. 

The role of this paper is to fufill the original goal of this project: 
to determine morphological classifications for galaxies within three 
clusters at $z=0.31$ and to analyse these in the context of hypotheses
discussed in Paper I and B96. The basic material for our analysis
consists of morphological classifications, uniformly constructed, for
magnitude-limited samples of cluster galaxies in AC~103, AC~118 and
AC~114, each of which exhibit a significant blue BO excess and are 
at $z=0.31$. At this redshift, we demonstrate that biases
in morphological classifications when using high quality WFPC-2 data
are minimal. We also present new spectroscopy extending considerably
the data published by CS and in Paper I and which enables us to analyse the
properties of a large sample of galaxies with $HST$ morphologies in the
context of spectral diagostics of recent star formation. This is the
most direct method to uncover the physical mechanisms which are causing
the recent and dramatic transformation in the cluster populations. Our $HST$  
cluster sample has a number of important advantages in this regard:  (1)\,the
three clusters lie at the same redshift and are imaged to comparable
depths -- comparisons between them are thus unaffected by differential
K-corrections or surface brightness dimming; (2)\,the
cluster masses and morphologies, as traced by their lensing and X-ray
properties (Smail et al 1997b, Natarajan et al. 1997), span a broad range  
and include a relatively low mass cluster (AC~103), a regular massive cluster
(AC~114), and a massive merging system (AC~118), allowing us to
differentiate between those triggering processes which depend upon
cluster environment; (3)\, finally, in the massive regular cluster
AC~114, we have obtained wide-field WFPC-2 imaging out to 
$\sim 600\,h^{-1}$ kpc (where $h$ is Hubble's Constant in units of 
100\,km\,s$^{-1}$\,Mpc$^{-1}$), enabling us to trace the morphological 
composition of this cluster outside the core and hence constrain those  
transformation mechanisms which are sensitive to local galaxy density. 
These are the first wide--field observations of the morphologies of 
galaxies in a distant cluster to be published and we show how they provide a
powerful insight into the nature and evolution of galaxies inhabiting such
structures.

The plan of the paper is as follows: In the next section we describe our
WFPC-2 observations and present the new ground--based spectra. 
In reporting the latter, we extend the sample of cluster members classified
according to their star formation activity by combining line index 
measurements made from the new spectra with existing broad--band colors. 
In \S 3, we review our scheme for morphologically classifying
galaxies both in terms of Hubble type and any visual evidence for dynamical
interactions. The results of our morphological classifications are presented
in \S 4 where we examine the overall morphological content of our clusters and
investigate the population gradients seen within our WFPC-2 fields. In
the next section we focus on the samples of spectroscopically confirmed
cluster members, in particular those involved in the starburst cycle. We
examine the morphology of galaxies in the different phases of this cycle
and in doing so review the importance of dynamical interactions. Our 
findings are then discussed in \S 6; in particular, we revisit the starburst  
cycle proposed by CS and explored more quantitatively by B96, addressing 
specifically whether one mechanism is sufficient to drive the phenomenon.  
We summarise our conclusions in \S 7. 

\section{Observations and data reduction}

\subsection{$HST$ Imagery}

WFPC-2 observations of AC~118, AC~103 and AC~114  were obtained over
the period September 1994 -- January 1996 in time allocated to two
different programs. The two clusters, AC~118 and AC~103, were observed in 
time allocated in Cycle 3 to complete the imaging of the 3 clusters studied 
by CS.  Due to scheduling delays, these observations were in the end carried  
over into Cycle 4, the first cycle after the HST refurbishment mission. The
other cluster, AC~114, was observed during Cycle 5 in time allocated
to Smail, Kneib \& Ellis to further investigate the gravitational
lensing of background sources by this cluster (Smail et al. 1995a, Natarajan 
et al. 1997). 

Details of these observations are given in Table 1 where we list the field
centers, the dates, the filters used, the telescope position angles (V3 axis), 
and the total exposure times.  All three clusters were imaged using the
F702W filter.  For AC~118 and AC~103, a single pointing at the listed position  
and PA was obtained, involving 3 orbit--length exposures offset from each 
other by 3 WFC pixels (0.30\,arcsec) to facilitate cosmic ray and hot pixel  
removal. The positions and orientations chosen for AC~103 and AC~118 were 
those which maximised our coverage of the existing samples of spectroscopically 
confirmed members (CS, Paper I) within the core regions of these clusters. 
Hence the centers of these clusters are not coincident with the apex of the  
WFPC-2 field. AC~114 was observed over 24 orbits, split into four different  
pointings, chosen to give both contiguous coverage of the cluster (over an 
area approximately $6.5' \times 3.9'$ in size) and a double exposure of the  
central core region of the cluster. The latter was achieved by positioning one  
of the WFC chips on exactly the same area in two of the pointings thereby 
giving a total exposure of 33.6\,ks on the central $1.3' \times 1.3'$ of the  
cluster. Yet again a 3 WFC pixel `dithering' between individual exposures was  
employed.  

Processing of our images took place using the standard STScI pipeline 
reduction, with the steps of bias subtraction, dark current removal and 
flat--fielding accomplished in the usual way. The images were then aligned 
using integer pixel shifts and combined with the IRAF/STSDAS task CRREJ. 
Cataloging and photometry of objects on these coadded frames was carried out 
using the SExtractor image analysis package (Bertin \& Arnouts 1996). Image 
detection proceeded by first convolving the data with a 0.3\,arcsec top--hat  
filter and then employing a $\mu_{702}=25.0$\,mag\,arcsec$^{-2}$ (equivalent to 
1.3$\sigma$ above the background) detection isophote together with a minimum 
object area of 12 contiguous pixels. The SExtractor package measures both
isophotal and total `Kron-type' magnitudes for each object and it is the 
latter that we use for this study. These have been placed on the 
$R_{702}$ system using the synthetic zero-points published by Holtzman et al. (1995).

The surface-brightness limits attained in our cluster exposures are
sufficiently deep to secure imagery with a signal-to-noise ratio high enough
to provide robust morphological classifications to at least $R_{702}\sim 22.5$
($\sim M^\ast + 3$\,mag), the limit achieved in the ground--based
photometric studies.   Our confidence in achieving this limit was based
on the analysis of the first `ERO' observations made with WFPC-2
immediately after refurbishment (eg. Dressler et al. 1994b), which revealed
the tremendous potential of HST for conducting morphological
classifications at these faint limits.  Such morphological
classifications are of course sensitive to both the bandpass and depth
of the source material used.  Our selection of the F702W filter
provides close to rest-frame $V$ imagery of the distant cluster members,
ideally matched to the blue/visual source material on which local
morphological schemes are based.  The issue of the depth required to 
adequately classify galaxies has been extensively addressed in the simulations  of Glazebrook et al. (1995) and Abraham et al. (1996). They investigated
the reliability of using F814W images of a few orbits duration 
for classifying field galaxies to $z\simeq$1. These workers found little
error in the recognition of spheroidal galaxies and early-type spirals, but
intermediate and late-type spirals at high redshifts appear to have
later Hubble types than might genuinely be the case because of a
combination of signal-to-noise ratio (S/N), sampling and bandpass shifting 
effects. However, our clusters lie at sufficiently low redshift
that we expect this bias in the spiral classifications to be
negligible.  For the case of differentiating between E and S0 galaxies
using WFPC-2 imaging we refer to the discussion in Ellis et al. (1997)
and Smail et al. (1997a).  These authors classified the
populations in $z\sim 0.55$ clusters using F814W images (also rest-frame $V$)
from WFPC-2. To a magnitude limit significantly fainter than that
adopted here, at a redshift where the scale and sampling is 30\%
coarser, they demonstrated that only $\sim 15$\% of the S0 population
was misclassified, the loss being round S0s misclassed as Es, a
problem which also occurs at a similar level in local samples
(c.f.\ Smail et al. 1997a).  We are therefore confident that the depth
and wavelength coverage available from our WFPC-2 imaging is sufficient
to provide robust and unbiased classifications on the Hubble system for
all morphological types in these clusters.  
  
\subsection{New ground--based spectroscopy}

Another important component of our observational program is the comprehensive 
spectroscopic study of our distant cluster fields. While at the 
most basic level spectroscopy is necessary to provide membership information 
for galaxies in our fields, we do not expect a high field contamination rate 
in our samples due to the high contrast of the cluster cores in the limited
field of view of WFPC-2 (Couch \& Newell 1984; hereafter CN).  Instead, the  
principal role of the spectral data is to obtain high quality line index
measurements for member galaxies, which we can combine with the detailed  
morphologies from our WFPC-2 imaging. Following CS and B96, we show that the  
information the former data provide on current and past star formation is  
crucial for understanding  the evolution of galaxies in these clusters. In 
obtaining new spectroscopy, we have sought to redress the incompleteness in 
CS's spectroscopic coverage to $R_{F}=20$ and, in a complete way, to extend 
it faintwards to limits more commensurate with the depth of our WFPC-2 images.

The new data presented here were obtained during two observing runs, one on the
3.9\,m Anglo--Australian Telescope (AAT) using the Low Dispersion Survey
Spectrograph (LDSS-1), and the other on the European Southern Observatory's
3.5\,m New Technology Telescope (NTT) using the European Multi--Mode 
Imager/Spectrograph (EMMI). For the former, the three clusters AC~118, AC~103, 
and AC~114 were targeted. Two multi--slit masks, each containing slits
for 20--25 objects, were designed and manufactured for each cluster. 
Program objects were chosen first and foremost from $K'\leq 18$ samples
assembled for each cluster from the infrared imaging data of B96. 
The purpose of this infrared ($K'$)--based selection was to reduce
the bias towards star--forming and, in particular, star--bursting galaxies, 
which as demonstrated by B96 is a serious effect in samples selected in the  
optical (rest-frame uv-blue). While this catered for object selection within 
the $2.8' \times 2.8'$ $K'$-band field and hence the $2.6' \times 2.6'$ WFPC-2 
field contained within it, it did not fully utilise the 10\,arcmin field of 
LDSS-1. Our samples were therefore supplemented with objects from the 
photographic catalogues of CN with a small number of the galaxies observed  
spectroscopically by CS also being included for comparison purposes. Selection 
in this case was made using optical data to a limit of $R_{F}\sim 21$. Of the  
$\sim$25 objects selected for each mask, it was possible to include 10--12  
within the single WFPC-2 field of AC~103 and AC~118, and all but 1 or 2 within  
the larger field of AC~114. These numbers were set largely by the need to 
impose a minimum slit length of 15\,arcsec to facilitate accurate sky 
estimation and subtraction.

For the NTT run, our observations were subsidiary to a program of spectroscopic
confirmation and redshift measurement of the gravitationally lensed `arcs' and 
multiple images in the field of AC~114 (Ebbels et al. 1997). Spectroscopy
of the lensed candidates was carried out using two masks, one designed to
cover the brightest arcs and another to cover a sample of fainter `arclets'. 
The unused regions of these masks were accordingly devoted to a set of `bright' 
and `faint' galaxy targets, respectively, within the AC~114 field. Selection of
the `bright' sample proceeded as for the LDSS-1 observations using
the $K'<18$ limited  infrared sample from B96 and the photographic $R_{F}$ 
photometry of CN. A total of 44 slits were devoted to
this sample, of which 25 lay on objects within our WFPC-2 field. For the
faint sample, object selection was carried out using the image
taken with EMMI for the purpose of mask design. A sample of 35 objects 
covering a similar magnitude range ($21.0\leq R \leq 22.5$) to the lensed 
candidates was chosen with 12 of these being within our WFPC-2 field.
    
The AAT run took place on the nights of 1995 August 28, 29 and 30. LDSS-1 was 
used with its ``high-dispersion'' grism to give a dispersion of
164\,\AA\,mm$^{-1}$ and, with a slit--width of 1.5\,arcsec employed, a spectral 
resolution of $\sim$9\,\AA. A thinned 1024~x~1024 24\,$\mu$m pixel Tektronix 
CCD was used as the detector, providing a wavelength coverage of 
4000\,\AA $< \lambda <$ 8000\,\AA. Excellent weather conditions were 
experienced on the first two nights of this run; the third night was 
completely lost due to cloud cover. 

The NTT observations were made on the nights of 1995 October 22, 23 
and 25. EMMI was used in its `Red Imaging Low Dispersion' (RILD) mode with
the 118\,\AA\,mm$^{-1}$ grism (grism \#2) and a thinned 2048~x~2048 
24\,$\mu$m pixel Tektronix CCD (\#36) as detector. A slit width of 1.5\,arcsec 
was used yielding a spectral resolution of 12\,\AA. The wavelength coverage of
the spectra was approximately 4000\,\AA $< \lambda <$ 8700\,\AA. All the 
observations were made in non--photometric conditions with the `faint' mask
exposures being particularly affected by cloud. 
  
A log of the AAT and NTT observations, with relevant details such as the 
total exposure time and the recorded seeing, is given in Table 2. 
The quoted exposure times consisted of a series of 2000\,s (and the
occasional 1000\,s) integrations for the LDSS-1 observations and 3600\,s 
integrations for the EMMI observations.

The data from both runs were reduced using a combination of routines in
the LEXT (Colless et al. 1990), FIGARO, and IRAF reduction packages. Individual
exposures were bias subtracted and then combined using the CCDPROC and 
IMCOMBINE commands within IRAF. Some of the LDSS-1 frames required small 
(1--3 pixel) shifts in the dispersion direction prior to coaddition to 
compensate for the effect of flexure within the spectrograph throughout the 
course of a set of integrations on a particular mask. The size of the shifts 
required were determined by cross-correlating the positions of the night-sky 
lines in adjacent exposures. The first night of data taken with LDSS-1 also
required correction for a small misalignment between the dispersion direction
and the columns of the CCD. This was accomplished using the SDIST and CDIST
routines within FIGARO. The former was used to quantify the tilt from the
spectra recorded in a tungsten lamp exposure taken through the same mask; CDIST
then applied the fractional shifts required in the spatial direction (ie. 
along each row of the CCD image) to align the spectra along the CCD columns.
The final steps of spectral extraction, sky-subtraction and
wavelength calibration were performed using LEXT as previously described in
Paper I. 

A representative collection of the reduced spectra is shown in Figure 1.
These spectra are comparable in quality to those obtained in earlier
work (CS, Paper I) with the same range in S/N (3--20) and a similar median 
value (S/N$\sim$8). They thus provide the same high rate of redshift  
identification and level of precision in making spectral line measurements. 
The only exception to this are the spectra 
obtained with the `faint' mask at the NTT. Apart from 2 objects, these spectra
all had signal-to-noise values too low (S/N$<3$) to make reliable absorption 
line measurements. Nevertheless, a large fraction of these objects did
provide redshift measurements due to the presence of emission lines.
Measurements of the central observed wavelength and equivalent width
(EW) of the identifiable emission and absorption lines in each spectrum were
carried out using the {\sc FIGARO} routine {\sc ABLINE}. This was done  
independently by two of us (W.J.C.\ and A.J.B.) in order to provide an 
empirical check on our errors. In particular it allowed us to assess the  
uncertainty introduced into our equivalent width measurements through the  
placement of the continuum -- a procedure which is done interactively within  {\sc ABLINE}.  

The results of these measurements together with other relevant data are 
presented in Table 3. The first 3 columns give the identity number of each 
object in (i)\,the SExtractor catalogue derived from our WFPC-2 images 
(\S 2.1), (ii)\,the photographic catalogue of CN, and (iii)\,the infrared 
($K'$) catalogue of B96, respectively. The absence of an entry in the 
first of these columns indicates that the object lies outside the WFPC-2 
field. Listed in the next 5 columns are the R.A.\ and Decl.\ (1950), the $R$ 
and $K'$ magnitudes, and the $B$--$R$ color. The $R$ and $K'$ magnitudes are  
`total' values, the former being taken from our WFPC-2 derived SExtractor  
photometry where available (and from CN where not) and the latter from B96. 
The colors
are the reddening--corrected $B_{J}$--$R_{F}$ values of CN, transformed
onto the standard Kron-Cousin $B$--$R$ scale (Couch 1981). The next 3 columns 
of Table 3 contain our spectral measurements. Listed are the mean redshift and 
the rest--frame equivalent widths of the [O\,II]$\lambda$3727 emission and 
H$\delta$ absorption lines measured for each object. The presence of a colon 
following the redshift value denotes that the measurement is based on just one 
emission line -- assumed in all cases to be [O\,II]$\lambda$3727. The mean 
internal error in our redshift measurements is $\Delta z = \pm 0.0003$
(corresponding to $\Delta v = \pm 70$\,km\,s$^{-1}$ in the rest-frame at $z=0.3$) and a comparison based on the 10 repeated CS objects indicates a mean  
redshift difference, $z$(this paper)--$z$(CS), of $-0.0001\pm 0.0004$ -- consistent with the internal errors in both sets of measurements.
The errors quoted for the [O\,II]$\lambda$3727 and H$\delta$ equivalent
widths represent the quadrature sum of the uncertainties arising from the
noise in the spectra (taken from the [S/N, $\epsilon$(EW)]--relation derived
empirically by CS from model synthetic spectra) and the placement of
the continuum (as determined from our two independent sets of measurements).

As a final step we combine our photometric data and spectroscopic measurements
to derive a spectral type for each object, the results of which are recorded
in the last column of Table 3. Here we follow exactly the precepts established
by CS based on the position of objects within the [EW(H$\delta$), 
($B$--$R)_{\rm obs}$]--diagram. As described in detail by CS and further
discussed in Paper I, this diagram is a powerful diagnostic for recognising
and therefore isolating galaxies with different ongoing and recent 
star-formation histories. In the context of the star formation activity seen
in distant cluster galaxies, 5 spectral types are of relevance:  
(1)\,``Starburst'' (SB) -- galaxies undergoing a burst of star formation, 
(2)\,``Post--starburst'' (PSG) -- galaxies which completed a burst of star  
formation within 2\,Gyr prior to the epoch of observation,  
(3)\,``Spiral--like'' (Sp) -- galaxies with the spectral and color 
properties  of normal nearby spiral galaxies, (4)\,``Passive'' 
(Psv\footnote{These were referred to as ``E/S0'' types in CS and Paper I;  
we have changed our nomenclature here to avoid confusion with our morphological
classifications (see \S 4).}) -- 
galaxies with the spectral  and color properties of nearby E and S0 galaxies, 
and (5)\,``Red H$\delta$--strong'' (HDS) -- galaxies with the same spectral 
and color properties as the passive population but with enhanced H$\delta$ 
absorption. The exact definitions of these different types and their location
within the [EW(H$\delta$), ($B$--$R)_{\rm obs}$]--plane can be found in CS 
and Paper I and these are used here with one exception: B96's more conservative 
limit of EW(H$\delta$)$>3.0$\,\AA\ for HDS types is adopted instead of the
EW(H$\delta$)$>2.0$\,\AA\ limit used previously. For AC~103, AC~118 and 
AC~114 we once again define ``red'' and ``blue'' galaxies to have 
$B$--$R$ colors (col.\ 8, Table 3) greater than and less or equal to 2.0,  
respectively, and stress that it is this color boundary which separates the 
HDS types (which are ``red'' and often referred to as ``E+A'' galaxies 
elsewhere in the literature; Dressler \& Gunn 1982, 1983) from the PSG galaxies  (which are ``blue''). 

In listing the spectral types for cluster members in Table 3, a number of
objects have been designated as ``Psv(pec)''. These are galaxies whose
spectra have all the same characteristics as the passive types but whose
colors are marginally bluer (ie. $B$--$R\sim$1.8--1.9) that they do not satisfy the `red' criterion. Whilst photometric scatter could account for 
some of these objects, we flag them nonetheless. 
It should also be noted that some objects have their spectral types listed as 
``?''. This indicates that their spectral types were not measurable due to 
the low quality of their spectra. 

These new observations have provided spectroscopy for 107 previously 
unobserved objects within the fields of AC~103, AC~118 and AC~114. Of these, 
15, 21 and 31 galaxies are contained within our WFPC-2 images of these clusters 
thereby extending the samples of spectroscopically confirmed cluster members in
these fields to 26, 30 and 73 galaxies, respectively. In addition, there is now
sufficient data to make comparisons between cluster and field objects with
the latter types amounting to a sample of 33 galaxies with a mean redshift
of $<\! z\! >=0.41\pm 0.20$. To this end we have compared the distribution in  
[OII]$\lambda$3727 equivalent widths to gauge whether there is any gross  difference in star-formation rates between the two environments at these 
earlier epochs. A Kolmogorov--Smirnov test shows that there is {\it no}  difference between the cluster and field distributions at the 5\% significance
level. Further spectroscopy of matched samples of cluster, super--cluster and  
field galaxies at these redshifts is really needed to establish this  
conclusively.

\section{Morphological Classification of Cluster Samples}

An inspection of our WFPC-2 images indicated that the goals of our
study could be best achieved by continuing with the visual morphological
classification scheme we developed for the WFPC-1 data in Paper I. 
Whilst machine--based measurements offer an objective means for obtaining
broad classifications across large samples (Abraham et al. 1996; Smail et al.  
1997a), the visual approach is more sensitive to finding regular galaxies
with unusual structure or minor peculiarities -- traits which we regard to be
more important to this study. The excellent quality of the WFPC-2 frames meant  
that they provided high signal--to--noise (S/N$>100$) imagery of {\it all} the  
galaxies whose evolution had been previously quantified both photometrically 
and spectroscopically. Most importantly, the S/N was clearly sufficient to
recognise small irregularities in their structure that might be pertinent to  
their evolution (eg.\ signatures of dynamical interaction). The primary 
concern of this section, therefore, is the visual classification of  
magnitude--limited samples of galaxies within our cluster fields. We also  
describe tests of our discrimination between E and S0 galaxies using a profile  
fitting technique.
 
\subsection{Visual Classification}

As was indicated in the initial $HST$ studies of distant clusters (Paper I; 
Dressler et al. 1994a,b; Oemler, Dressler \& Butcher 1997) and which is 
even more striking in post-refurbishment 
images, the galaxy populations within these fields appear, morphologically,  
to be predominantly of normal Hubble type (elliptical, S0, spiral). In 
addition, the WFPC-2 data continue to reveal a significant number of objects
with conspicuous signs of dynamical interaction either through the presence
of tidal bridges and tails or gross distortions. We have therefore continued 
to use the classification scheme developed in Paper I based largely on the 
Revised Hubble system with an additional classifier used to flag different 
manifestations of interactions and merging between galaxies. Broadly, we
assign to each object a Hubble type (ie. E,S0,Sa,Sab...Irr) and categorise 
dynamical interactions on the basis of the visible evidence be it an obvious
merging of two objects, a clear interaction between two objects, or the 
presence of tidal features. The presence of satellites is also noted. The 
criteria used in making these classifications will be detailed below. 

Two of us (W.J.C.\ \& R.S.E.) each independently classified galaxies in all
our cluster fields in the following manner. Using AC~103 as a test case, 
all objects within the SExtractor catalogue were classified in terms of
Hubble type to as faint a magnitude limit as possible ($R_{702}\sim 23.5$). 
We then
determined how the incompleteness of our classifications varied with 
apparent ($R_{702}$) magnitude. This was gauged by the numbers of objects which
were either too faint to classify or when classified as a spiral, their
spiral arm structure was too faint to determine their subclass (i.e.\ a,b,c,dm).
It was found that in this case of a 3--orbit exposure, incompleteness began 
to set in at $R_{702}\sim 22$, rising from $\sim 5$\% at $R_{702}=22$ to 
more than 
20\% at $R_{702}=23$. This led us to adopt a magnitude limit of 
$R_{702}=22.25$ ($M_V \sim M^\ast + 3$ at $z=0.3$) for our morphological  
classifications in AC~103 and AC~118. The corresponding limit adopted for 
the 6--orbit exposures of AC~114 was $R_{702}=23.00$\,mag 
($M_V \sim M^\ast + 3.5$). 

A full classification down to these magnitude limits was then conducted in
each of our cluster fields. The two sets of results were merged and the final 
morphologies are presented in the last 4 columns of Table 4. Objects with 
stellar appearance have been omitted in the compilation of this 
table. The first of these columns contains the Revised Hubble type. Here the 
nomenclature used is standard with the only departure being for the brightest 
member of AC~114 whose visual appearance was more appropriately described by 
the ``cD'' classification of the Morgan system. In denoting early Hubble types, 
the classifications ``E/S0'' and ``S0/a'' have sometimes been used. These 
reflect the ambiguities and uncertainties that can occur in classifying these 
types (this is discussed more fully in Smail et al. 1997a) and are not meant 
to indicate transition cases. More specifically, they have been used to record 
cases where either both classifiers were unable to decide between an 
E or S0/S0 or Sa classification or their classifications differed to this 
extent. The relative split between these two different situations was  
approximately 60\% : 40\%. Spirals have been classified to half a subclass 
with barred systems denoted in the usual way. However, barred systems were only 
noted where obvious and in no way can our catalogues be regarded as complete in 
the recording of these types. As partly mentioned already, some objects could 
be readily identified as ``spirals'' (through their ``disky'' rather than  
spheroidal appearance) but as a result of their faintness, orientation, or 
proximity to another object, the subclass they belonged to was unclear. Such 
cases are denoted with an ``S''. 

A small fraction of objects have a morphology that could not be described 
within the Revised Hubble system; these are listed as either peculiar (``pec'') 
or compact (``comp''). The latter, which were the most common in this category, 
are an interesting set of barely resolved objects which while not truly  
stellar in appearence, are too compact to classify in any  detail. 
Unfortunately no spectroscopy is yet available to cast any light on the nature  
of these objects. Finally, some objects have been given a ``GL'' 
classification; this category is used for the numerous gravitationally lensed  
``arc'' features revealed in our $HST$ images (Smail et al. 1995a). Our list, 
however, should not be taken to give a complete identification
of such images with a more optimised and thorough compilation being reported 
elsewhere (Smail et al. 1997b; Natarajan et al. 1997). There is also a small 
number of galaxies which are well enough resolved to be given a Hubble  
classification but are clearly distorted in a manner consistent with being  
lensed by the cluster. Such cases have been catalogued with their Hubble type  
and a note made to this effect in the last column of the table. 

The next column, headed ``MDS'', lists our classifications on the numerical  
scheme devised by Glazebrook et al. (1995) in their analysis of the $HST$  
Medium Deep Survey data. Here the different Hubble classes are assigned a 
number between 0 and 6 with E=0, E/S0=1, S0=2, Sa--Sb=3, S=4, Sc--Sdm=5, and  
Irr=6. This scheme then uses a value of -1 for objects with the same 
``compact'' morphology as found here and a value of 7 for unclassified objects.  The only convention not adhered to was to assign galaxy mergers a value of 8.  
Instead we have preferred to record the MDS number appropriate to the Hubble  
type of the object and to use the subsequent column to indicate its involvement  in a merger. 

The third of our classification columns in Table 4 is used to indicate 
evidence of dynamical interactions and/or the presence of satellites. The 
morphological signatures associated with interacting and merging systems are 
well documented and understood from the observation and modelling of nearby 
systems (Arp 1966, Toomre 1977, Quinn et al. 1993), and the simulations of 
Mihos (1995) provide a useful indication of how the different phases of the 
merging process might appear in WFPC-2 images of the distant galaxies studied 
here. With these studies in mind, we have categorised objects in terms of three 
generic situations:

	1. {\it Interactions}.  These are generally the most conspicuous and 
hence easily identifiable cases involving spatially distinct systems 
(separations $\geq 0.5$\,arcsec) of comparable brightness which have either a 
clear isophotal and hence tidal link between them (eg. AC~114\#CN87 and 
AC~114\#CN849 in Fig.~6a) or are mutually distorted in a way consistent with  
tidal interaction (eg. AC~114\#CN111 in Fig.~6a).\\
\indent 2. {\it Mergers}.  These fall into two categories: multiple systems 
clearly in the process of merging and single systems where the visible 
distortions/peculiarities are interpreted as being due to a recent merger. The 
former are distinguished from objects in the previous class in that their 
multiple structure is on smaller scales (ie. separations 
$\sim$0.2--0.5\,arcsec; $<3.0\,h^{-1}$\,kpc at $z=0.3$) -- examples include  
AC~118\#CN104 and AC~114\#CN22 in Fig.~6a. The latter systems are identified 
on the basis of either having the classical tidal arm signature of a merger  
event (eg.\ Fig.~1 of Mihos 1995) 
or being so strongly distorted as to suggest the recent capture of another  
galaxy. Obviously the identification of these types is a rather subjective 
procedure and we treat such cases with caution (see below). A typical example
is AC~114\#CN155 in Fig.~6b.\\
\indent 3. {\it Tidal Evidence}.  This category is used to record objects which 
show evidence of tidal features but do not provide direct evidence of a
merger or interaction with another galaxy. The connection between the two
is, therefore, by implication only. Such features include: possible ``shells'' 
of the type seen by Malin \& Carter (1983) in the outer regions of nearby 
spheroidal galaxies, warped and distorted outer isophotes around elliptical
galaxies, ``debris'' of the sort predicted by Mihos (1995) to surround merger
products for $\sim 1$\,Gyr after the merger event\footnote{A preliminary 
determination of the incidence of these types in AC~103 and AC~118 was
reported in  B96.}, and faint tails, whisps and arms consistent with having
a tidal origin. 

These three different classes are denoted I or i, M or m, and T or t in the
``Dyn/Int'' column. The use of upper and lower case symbols is
to indicate the level of certainty and unanimity in our classifications. 
Where upper case symbols are used, both classifiers independently and 
unreservedly gave the object this classification. Lower case symbols then
indicate instances where there was not unanimity or some doubt in the 
classification was expressed by either or both of the classifiers. 
The one other entry in this column is ``Sat'' for galaxies with faint
satellites. The criterion applied here is the presence of significantly
fainter objects (as determined only visually) {\it within a galaxy's visible 
envelope}. In cases where there are multiple satellites, the exact number is 
recorded in parentheses.

The final column in Table 4 contains comments relating to our 
morphological classifications. These serve several purposes, the main two
being to make it clear what type of feature is responsible for an object's
T/t classification, and to record peculiarities which are not described
adequately under the Hubble or I/M/T/Sat schemes. One such peculiarity 
commonly recorded was a clear asymmetry in the arm structure of spirals.

Having presented our morphological classifications, some comment is required
on their reliability and consistency both internally and externally. An
internal check was available from the two independent sets of classifications
and this showed a high level of concordance between the two classifiers with no
evidence of any systematic difference between them. In particular, a comparison 
of the Revised Hubble types derived over the three cluster fields indicated 
that, on average, exact agreement between the two classifiers occurred 
56\% of the time, agreement to within half a class (i.e. the difference
between an E and E/S0 or between an Sb and Sbc) occurred 86\% of the time, 
and agreement to within one class occurred 95\% of the time. In addition, the 
average of these differences was zero indicating that one classifier 
did not systematically classify objects earlier or later (in the Hubble sense)
than the other. When differences did occur, the method for dealing with them
was as follows: In cases where the difference was one class or less, the
average between the two was taken. For the small number of more
serious discrepancies, the two classifiers reviewed their classifications 
and a final result was mutually agreed upon.  

The excellent level of internal agreement of our classifications, when
combined with our concious attempt to keep our classification scheme as
close as possible to the systems used by other HST morphological
studies, means that our catalogues can be easily compared to these
other samples.  In particular, we can be confident that our Revised
Hubble types are on the same system as those derived visually for the
field galaxy populations imaged in the HST Medium Deep Survey (MDS,
Griffiths et al. 1994; Glazebrook et al. 1995), as a large part of
their classifications was conducted by R.S.E.  In addition, both
W.J.C.\ and R.S.E.\ played a major role in a similar classification of
galaxies in more distant clusters as part of the ``MORPHS'' project --
an HST--based study of 10 clusters at $0.37\leq z\leq 0.56$ 
(Smail et al. 1997a). This ensures that our classifications are well tied 
to this study as well.

Finally, as a reference for examining how the different morphological types
we have identified are distributed spatially within the clusters (\S 4.4), plots
showing the location and morphology of the classified samples within the
WFPC-2 fields are presented in Figure 2. Galaxies have been plotted using 
different symbols to differentiate their morphology and those with a M, I, or 
T classification have been circled. 

\subsection{Profile fitting}

An important check on the accuracy of the visual
morphological classification of E and S0 
galaxies can be made using
surface brightness profiles to identify extended
disk components. The method we have adopted is
to visually classify the surface
brightness profiles, using plots of
$\mu$ vs $r^{1/4}$, $\mu$ vs $r$ and $C_4$ vs $r$, where
$r$ is the equivalent radius of the best-fitting elliptical
isophote with surface brightness $\mu$ in logarithmic
units and $C_4$ is the $\cos4\theta$
coefficient in the Fourier decomposition
of the fit. In the absence of instrumental PSF and finite pixel
size effects, a pure de Vaucouleurs $r^{1/4}$ elliptical would
therefore be a straight line in the former plot, and an
exponential disk in an S0 would show up as a linear
region in the outer parts of the second plot. The final plot
measures the `diskiness' of the isophotes and
is a useful diagnostic for the presence of nearly edge-on
flattened disks (e.g. Carter 1987).

The surface brightness profiles were extracted from the
reduced HST images using the Starlink Extended Surface Photometry
package (Privett 1996) which allows interactive fitting
of the local sky background and galaxy profiling
using elliptical isophotes fitted to uncontaminated
parts of the image. To account for the effects of
finite pixel size and the wings of the WFPC-2 PSF 
on the profiles, we used the IRAF ARTDATA package
to generate synthetic images using pure exponential
and pure $r^{1/4}$ profiles, with appropriate S/N,
and convolved them with PSFs generated using the
TINYTIM (Krist 1994) package. These profiles then served
as a reference set on which to base classification of
the profiles extracted for galaxies in the 3 clusters.

The primary role of the profile-based classifications
has been to resolve cases of ambiguity where the 
visually determined morphologies were discrepant between the two
classifiers. Figure 3 shows two such examples
from AC~103 where the profiles indicate an elliptical (HST\#69)
and S0 (HST\#650) classification, respectively. We have also
used blind profile classifications to check the 
morphological classifications for a subset of the 
E \& S0 galaxies in each cluster.
The profile classifications were found to 
agree exactly with the visual
classifications in 70\% of cases; 15\% of the sample
were classified E visually but S0 on the basis
of profiles, whilst 15\% were classified S0 visually 
and E from the profiles (3\% of the sample 
were unclassifiable on the basis of profiles alone).
In general, therefore, despite using
a very simplified model of real galaxies, the profile
classifications support those based on a direct visual inspection 
of the HST images. In cases where there is disagreement, the profile--based 
classifications have been adopted in preference to the visual ones.  

\section{Results}

\subsection{Morphological Content}

Determining the morphological composition of our distant clusters is the
primary goal of our study and, as a first step, we examine the distribution
in Hubble type of the magnitude--limited samples classified in each field.
This information is contained within Figure 4 where we present ``cluster''
morphological distributions which show the numbers of different Hubble
types (binned into E, S0, Sa--Sb, Sc--Sdm, Irr) corrected statistically
for the expected contributions from the superimposed field galaxy population.
The raw numbers of objects classified as ``pec'' or ``comp'' are also shown.
Because of the much larger areal coverage available for AC~114, the data 
for this cluster have been divided into 3 contiguous radial bins each 
1\,arcmin (180\,$h^{-1}$\,kpc) in width and plotted accordingly. A radial
analysis of the data is justified for this cluster given its regular, 
centrally--concentrated structural morphology (CN). Note also that the 
inner $0\leq r\leq 180\,h^{-1}$\,kpc zone corresponds closely to the 
regions covered by our single WFPC-2 pointings of AC~103 and AC~118. 

Subtraction of the field component from our cluster distributions was 
accomplished using the $r_{\rm F}$--band galaxy number count data of 
Metcalfe et al. (1995) and the deep $HST$ field morphology data of 
Driver et al. (1995). The former were used to determine the expected number of 
field galaxies within our WFPC-2 frames; they predict 27 galaxies brighter 
than $R_{702}=22.25$ within the single 5.1\,arcmin$^{2}$ WFPC-2 fields 
acquired for AC~103 and AC~118, and 195 galaxies brighter than 
$R_{702}=23.00$ in the 20.1\,arcmin$^{2}$ field of AC~114. In comparison, 
the numbers of galaxies observed in these fields to the respective limits are
AC~103 (122), AC~118 (129), and AC~114 (474). The distribution 
of the field population across the different Hubble types was set using the  
morphological mixes published by Driver et al.\ For our $R_{702}\leq 22.25$ 
samples, a \%E:\%S0:\%Sab:\%Scdm:\%Irr = 17:17:25:30:11 mix was adopted, and 
for our $R_{702}\leq 23.00$ sample in AC~114 a 
\%E:\%S0:\%Sab:\%Scdm:\%Irr = 10:10:25:30:25 mix was adopted. 

In determining the field contribution, careful account was taken of the 
differences between the $R_{702}$ system used for our WFPC-2 based photometry
and the $r_{\rm F}$ ($\cong R$; Couch \& Newell 1982) system used by 
Metcalfe et al. This was done by computing the offsets between the two bands  
using the spectral energy distributions for different Hubble types published by 
Pence (1976): $(R_{702}-r_{\rm F})=-0.22$ for E/S0, 
$(R_{702}-r_{\rm F})=-0.17$ for Sab, $(R_{702}-r_{\rm F})=-0.10$ for Sbc, 
$(R_{702}-r_{\rm F})=-0.08$ for Scd, $(R_{702}-r_{\rm F})=-0.06$ for Sdm.
It should also be pointed out that the field galaxy mixes used here were  
determined from WFPC-2 observations made in a different
passband (F814W$\approx I$) to the F702W band used for this study. However,
the difference in apparent morphological mix between these two red bands
should, to the limits we are working at, be small, to the extent that they
will have a negligible effect on our final field--subtracted ``cluster''
distributions. To be consistent at least in magnitude limit, we converted our
$R_{702}$ values onto the $I_{814}$ system assuming $R_{702}$--$I_{814}=0.3$ -- 
the mean color of the field galaxy population at these limits (Smail et al.  
1995b). 

To provide a present-day ``bench mark'' with which to compare the morphological 
composition within our distant clusters, predictions of the morphological mix  
based on that observed in similar environments nearby were computed in each
case. This is important in isolating evolutionary changes given 
that morphological mix is a strong function of global cluster morphology 
(Oemler 1974) and local galaxy density (Dressler 1980), both of which vary
considerably over our distant sample. To produce quantitative predictions, 
we used Dressler's well-defined 
[morphological mix, local galaxy density]--relation which he showed underlies
the variation in morphological mix with cluster morphology. Our procedure
was to evaluate the density of galaxies local to each of the
objects classified in our magnitude--limited samples, following as
closely as possible the prescription adopted by Dressler. This involves
identifying the 10 nearest neighbouring galaxies and then calculating
the corresponding projected surface density, corrected for field galaxy
contamination. In Dressler's nearby study this was done in the visual
band to a limiting magnitude of $m_{V}=16.5$ for a cluster sample with
an average redshift of $z=0.04$. At $z=0.31$ these experimental
``conditions'' can be reproduced almost directly, since in the
rest--frame our F702W band is reasonably well matched to Dressler's visual
band. The appropriate limiting magnitude in the $R_{\rm F}$ band at
this redshift has already been calculated by Couch (1981) and so we use 
the equivalent limit on the $R_{702}$ system ($R_{702}^{Dressler}=21.08$)
as determined by the transformations described above.

The distant versus nearby comparison of morphological content has been 
analysed in two ways. In AC~114 where our WFPC-2 imagery goes well out beyond
the central core region, the observed field--corrected E:S0:Sp mix has been
evaluated in contiguous radial zones centred on the central cD galaxy 
(HST\#1.1089) and 0.5\,arcmin ($90\,h^{-1}$\,kpc) in width. This is a valid
approach for reasons already mentioned. The local galaxy density averaged over 
all the galaxies within each of these zones was calculated, thus providing
a present--day comparison by reading off the corresponding E:S0:Sp mix
from Dressler's [morphological mix, local galaxy density]--relation.
In AC~103 and AC~118, where we have only single pointings with
the clusters not perfectly centred within them, such a radial analysis
was not practical. Instead, we simply compared the morphological mix observed
over the whole frame with the nearby prediction based on the local
galaxy density averaged over the whole frame as well.

\subsection{The Morphological Mix in the Cluster Cores}

We start by discussing the morphological mix seen in the cores ($r\ls
200\,h^{-1}$ kpc) of these clusters. Here the distributions in Figure 4 show
quite clearly that, with the exception of the ``Irr'' class, {\it all} Hubble
types are present in significant numbers in these regions. In particular, we
see considerable numbers of Sa--Sdm galaxies in addition to the dominant 
population of E and S0 types. This corroborates the conclusion of the initial  
$HST$ studies (Paper 1; Dressler et al. 1994a,b; Oemler et al. 1997)
that {\it spiral galaxies are a common inhabitant in the core regions of rich  
clusters at these earlier epochs.} Importantly, this mix of morphologies
is also seen in the samples of spectroscopically confirmed members 
(represented by the cross--hatched histograms in Figure 4) thereby ruling
out that it is an artifact of incorrect field galaxy subtraction. 

These trends are documented more quantitatively in Table 5 where we
list the E:S0:Sp fractions observed in our clusters together with those  
predicted from the Dressler 
[morphological mix, local galaxy density]--relation.  To aid our comparisons, 
we include for AC~114 the observed and predicted mixes  
within a single WFPC-2 sized field, centred on the cluster in 
an equivalent way to that of AC~103 and AC~118 (referred to as the ``WFPC-2''
region in Table 5). Inspection of the data shows only one instance where the 
observed and predicted mixes come within close agreement: in the central 
$\sim 1.0$\,arcmin ($\sim 180\,h^{-1}$\,kpc) in AC~114. {\it Clearly the 
process(es) responsible for the morphology--density relation have almost run  
their full course in the very centre of this cluster and it maybe no 
coincidence that this happens to be the most massive and regular cluster 
within our sample.} In the core regions of the other two clusters spirals are  
seen in excess numbers, making up $\sim20$--35\% of the galaxy population in  
contrast to fractions of less than 10\% seen at the same projected densities  
nearby. The highest fraction (35\%) is observed in AC~103, the least massive 
of the clusters in our sample. Furthermore, the enhancement in spiral numbers 
persists all the way into the centres of AC~103 and AC~118, with the E:S0:Sp  
mixes within their central 0.5\,arcmin being no different to those  
evaluated over the full WFPC-2 field. With spirals being present in excess
numbers, accordingly there is a deficit in the numbers of E and/or S0  
galaxies, the balance of which appears to vary between the two clusters. 
In AC~103, both the E and S0 fractions are below the present-day levels 
whereas in AC~118 it is only the ellipticals that are deficient. 

A comparison of the morphological mixes we see at $z=0.31$ with those
at higher redshift is also an important one and here the ``MORPHS'' 
observations based on single WFPC-2 pointings of the cores of 9 clusters at 
$0.37\leq z\leq 0.56$ (Smail et al. 1997a) allow us to do so almost directly. 
The two most conspicuous trends involve the Sp and S0 populations. The
spiral fractions in the higher redshift sample are, in general, significantly
higher than that seen at $z=0.31$, being at the 40--50\% level. In addition, 
we see a significant drop in the fraction of S0's in going to higher 
redshift; whereas $\sim 40$\% of the galaxies are S0s in our clusters at  
$z=0.31$, the fraction is consistently within the 5--20\% range in the
interval $0.37\leq z\leq 0.56$. Given that the morphological classifications
of both studies should be consistent with each other (see \S 3.1), it is 
unlikely that this is the result of systematic differences in the way
S0s are identified. Rather it further emphasises the conclusion of 
Dressler et al. (1997a) that there has been a significant and quite rapid 
production of S0s (and depletion of Sps) in cluster cores since $z\sim 0.5$.

\subsection{The morphological mix on larger scales}

Our WFPC-2 mosaic of AC~114  provides a first--time view of the 
morphological mix in the more outer regions of a distant cluster; 
in particular, we are able to examine morphologies out to distances of
$\sim 600\,h^{-1}$\,kpc from the center of AC~114. In doing so, we probe 
regions where the projected galaxy density is a factor of $\sim 3$ lower than
that in the central core and, as a result of AC~114's rather flat and
regular surface density profile at these projected radii, is relatively uniform 
(50--60\,gals\,Mpc$^{-2}$).

Returning to Figure 4, the most conspicuous trend seen in going to larger
radii is the significant decline in E \& S0 numbers to the point that 
the distributions are now spiral--dominated. The extent to which this simply
reflects the change in morphological mix with galaxy density, which drops 
with radius, requires examination and here we refer to the data presented 
in Table 5 and Figure 5. The latter shows, in separate panels,  
the observed and predicted fractions of E, S0 and Sp galaxies in the 6  
contiguous radial zones evaluated for AC~114. The two innermost ``core''  
zones show graphically the behaviour described in the previous section.
 
At larger radii ($r>200\,h^{-1}$\,kpc) the discrepancy between the observed
and present-day fractions becomes very pronounced.  
This is particularly so for the spirals which show a {\it very strong 
increase}, rising to $\sim 60$\% of the cluster population which is well 
above the $\sim 20$\% levels seen nearby at the same, lower projected galaxy 
densities. This excess of spirals is largely at the expense of the S0 
population which shows a significant deficit at these larger radii. A deficit
in the fraction of E galaxies is also seen although it is only significant
at radii between 200 and $300\,h^{-1}$\,kpc. 

Some caution needs to be taken in interpreting the excesses
and deficits we have identified via this comparison. Firstly, the 
present--day mixes taken from Dressler's [morphological mix, local galaxy  
density]--relation represent {\it mean} values about which he found a scatter 
of $\pm 5$\%  from cluster to cluster. Secondly, the interpretation that 
the E and S0 deficits seen in our distant clusters are evidence for further 
production of these types at subsequent epochs is, with our use of fractional
content, not a unique one. A mere drop in the numbers of spirals would 
be sufficient to cause the E and S0 fractions to rise. Such a scenario cannot
be overlooked given the considerable ($\sim 2$\,mag) fading spirals are 
known to undergo should their star formation cease (B96, Abraham et al. 1996),  
thereby effectively removing many of them from apparent 
magnitude--limited samples such as we have studied here. The possibility that  
the distant cluster spiral populations are being modified in this way will be  
discussed in \S 6.   
 
\subsection{Trends in other morphological signatures}

As a final step in examining our observed morphologies, we focus on the 
subset of objects showing evidence of dynamical interactions, containing
satellites, and exhibiting structural abnormalities. In particular, 
we examine whether they show any spatial gradients or segregation within 
the regions covered by our WFPC-2 imagery. Any variations of this nature may 
further constrain what physical processes are at play; for example, cluster tidal forces (Byrd \& Valtonen 1990, Valluri 1993) which increasingly 
distort galaxies as they come close to the center, the process of 
violent relaxation which may mix the cluster uniformly but also destroy
interacting systems (Gunn \& Gott 1972), or ``galaxy harassment'' where a  
combination of two--body and tidal forces combine to disrupt low surface
brightness galaxies and those with low central concentration (Moore et al. 
1996). 

A visual assessment of the spatial distribution of dynamically interacting
(M/I/T) systems within our cluster fields is easily gained from the 
positional/morphological maps of Figure 2. Here we see no evidence for
any strong clustering of these objects nor any evidence that they avoid 
the densest regions of the clusters. At most there is an indication in 
AC~103 and AC~118 that they are located preferentially near the peaks in
galaxy density -- top right-- and bottom left--hand corners of
the AC~103 field and left/below center in AC~118 -- but it is hardly 
convincing. To be more quantitative, we have analysed the
number of M, I or T types (as a fraction of the total cluster population) as
a function of radius from the cluster center. The data for AC~114 are shown in
the bottom panel of Figure 5; this shows the fraction of M/I/T systems remains
relatively constant at $\sim 20$\% with the fluctuations about this being no
more than that expected statistically (Gehrels 1986). 
Although not plotted, AC~118 shows a similar flat behaviour out to its 
radial limit of $r\sim 300\,h^{-1}$\,kpc, albeit about a lower mean fraction 
of 12\%. The same analysis was not attempted for AC~103 due to its lack of  
adequate statistics and its irregular structure.

The statistics on those systems showing structural abnormalities or
containing satellites are small and so only broad conclusions can be drawn
about their distribution within the clusters. With the exception of one or
two cases, all the identifications of structural abnormalities occurred 
amongst spirals, their arm structure having been listed as either 
``assymetric'' or ``distorted''. We therefore compared the incidence
of these abnormal spirals in and out of the core regions in AC~114 and AC~118, 
with the contamination by field spirals taken into account. In AC~114, there
is no evidence to suggest that such systems are any more abundant in one
region than the other; about 1 in every 5 cluster spirals appears to have
this peculiar arm structure right across the cluster (0--0.5\,Mpc).
In AC~118, there is marginal evidence that the incidence of these galaxies
is greater outside the core but the factor of $\sim 2$ increase is only
significant at the 2$\sigma$ level. 

For those galaxies identified as having one or more satellites, a total of 
9 were found in AC~103, 7 in AC~114, and 5 in AC~118. Yet again we 
see cluster-to-cluster differences in their distribution rather than the
same general trend across all 3 clusters. In AC~118 it is notable that all
the galaxies with satellites are located within the central 1.0\,arcmin 
core whereas in AC~103 and AC~114 they show no such preference. These
observations must be treated with caution, however, as our cataloguing of
these systems is not in any way complete.

\section{Morphology and Star Formation}

Spectroscopy of the clusters studied here, as indeed with other distant 
clusters, has linked the BO--effect to higher levels of star formation, much 
of which is of an intense and short-lived nature (Dressler \& Gunn 1982, 1983;  
CS). We now address the morphological nature of these highly active systems 
on a case by case basis. This follows on from Paper I where we first examined 
the connections between star formation characteristics and morphology. However, 
on that occasion we were frustrated by the rather small sample of objects at
our disposal, particularly active blue members. With our sample now 
considerably expanded both in terms of the number of clusters imaged with 
$HST$ and the numbers of spectroscopically confirmed members with reliable 
star-formation typings, we are in a position to draw much firmer conclusions 
on this issue.

In Figure 6 the WFPC-2 images of the spectroscopically identified `active' 
members in AC~103, AC~118 and AC~114 are assembled and grouped according to  
their star formation typings (SB, PSG, Sp, HDS; see \S 2.2). The SB and PSG 
galaxies are shown in Fig. 6(a). The former, which occupy the top row, are
objects genuinely undergoing a major burst of star formation with 
high-excitation emission line-dominated spectra similar to local examples
of starburst galaxies and rest-frame colors equally as blue (see CS). 
The PSG objects which follow, are displayed in order of increasing 
$t_{\rm SF}$, the time that has elapsed since the completion of their most
recent star formation. This can be estimated from their position in 
the [EW(H$\delta$), ($B$-$R$)]-diagram using the model tracks of CS and B96.  
Going from left to right across each row and down the  mosaic, the first 7 
objects are estimated to be seen at $t_{\rm SF}\ls 0.4$\,Gyr, the next 4 at 
$t_{\rm SF}\sim 1$\,Gyr, and the final object at $t_{\rm SF}\sim 2$\,Gyr. 
Of these, the objects shown in the second row of Fig. 6(a) are the most
unambiguous in terms of having a starburst event in their recent past; their
colors are simply too blue [$(B-R)_{\rm obs}\sim 1.2$] and their H$\delta$ equivalent widths too large ($>7$\AA) to be consistent with the truncation
of star formation in just a normal spiral galaxy. For the remaining objects 
it is not so easy to distinguish them from ``truncated spirals'' although 
in the cases of AC~114\#228, AC~118\#166, AC~114\#191, and AC~114\#849, their
positioning very close to or below the spiral galaxy sequence in the  
[EW(H$\delta$), ($B$-$R$)]--plane favours more a starburst history (see CS
for further details). 

The most conspicuous feature of this set of objects is the number which  
are unambiguously interacting and/or merging, with 2 of the
3 SB galaxies and 3 of the 10 PSG galaxies having an I or M classification. 
Furthermore, they are all {\it major} mergers/interactions in which the 
participating galaxies are both luminous and of similar size. Our WFPC-2 
images of CN\#111, 87, and 22 in AC~114, of course, reconfirm the 
interacting/merging nature of these objects first reported in Paper I. 
However, we now see much more clearly the morphology of the individual 
galaxies that are involved. What is even more significant is that all but
two of the 6 galaxies for which we have the clearest spectroscopic and 
photometric evidence of starburst activity (top two rows of Fig. 6a) are  
undergoing major mergers/interactions: {\it Although our sample of such 
objects is small, it provides a strong indication that dynamical interactions
are a major driver of the starburst activity seen in these clusters.} Our  
observations also fit nicely with the sequence of events predicted by the  
simulations whereby the very active period of star formation triggered by the interaction/merger occurs close to the time we see them here of final
approach and coalescence (Mihos 1995). In addition, the galaxies involved 
are mid- to late-type spirals, consistent with the gas--rich systems 
required to sustain such a strong burst of star formation (CS). 

Whilst having established this important link between interactions/mergers and 
starburst activity, it clearly is not unique: we see starburst objects where 
major interactions/mergers could not possibly be responsible for their
star formation activity. The two cases in point are AC~114\#510 and
AC~103\#132. The former, which is caught undergoing a starburst, has the
morphology of a normal mid--type (Sb) spiral galaxy, albeit with a noticeably  higher central surface brightness. It is perhaps also of note that this galaxy 
is considerably further from the cluster center than the other two SB galaxies
(500\,kpc cf. 100--200\,kpc). AC~103\#132 is a late--type Scd spiral with
somewhat peculiar internal structure. One interpretation of this structure
is that it has undergone a minor merger but the evidence is hardly convincing 
(hence the ``m'' classification). 

A further conspicuous division seen amongst the galaxies of Fig.6(a) is the
common and quite distinct morphology of the bottom set of 7 objects, with 
all but one of them being normal, isolated S0--Sb types. Whatever the exact  
details of recent star formation are in these galaxies (ie. starburst, 
truncated spiral, or otherwise), clearly the end-products are early--type 
{\it disk} systems. While this identifies one evolutionary path at least that  
leads to the production of S0s, it raises other questions. For example, can 
any of these objects really be the descendants of the merging/interacting--type 
systems seen in the top two rows of Fig. 6(a)? We shall defer discussion of 
this issue until the next section.  

We next move on to the sample of 8 ``Sp'' types which have spectra and  
broad--band colors similar to present--day spiral galaxies; their images are  
presented in Figure 6(b). They show CS's spectral/color--based classification 
to be 100\% successful in identifying galaxies which are true spirals in the morphological sense. The fact that 7 of these spiral members are  
of Sb Hubble type or later further emphasises the late-type composition of the  
blue spiral population in these distant clusters (see also Oemler et al. 1997).
Our $HST$ images also convey a strong visual impression of what feeble and 
low surface brightness objects some of these galaxies -- particularly the 
later types -- would be in the event of their star formation ceasing and the 
associated bright knotty regions being no longer present. That some of these  
spirals will simply fade from view when the star formation in their disks is
eventually extinguished is clearly a distinct possibility. It is also of 
note that this set of galaxies is not without its irregularities. 
AC~114\#243 and AC~103\#134 are clearly interacting with or tidally distorted 
by their close neighbours while AC~114\#155 has tidal arm structure highly  
suggestive of a recent merger. We would stress, however, that galaxies in this  
class do not share the more definitive merging characteristics of the SB and 
PSG types. As already discussed, another important irregularity seen in many of 
the spirals is a strong asymmetry in their spiral arm structure (see \S 4.4). 
Such a peculiarity in the detailed structure of spirals in the rich cluster  
environment at these redshifts appears to be common; a more rigorous 
quantification of this effect based on a much larger sample of 
spectroscopically confirmed cluster spirals can be found in 
Dressler et al. (1997b).

Figure 6(c) contains the images of the red HDS types. It can be seen that 
the large majority (14/16) of these objects are normal, mostly isolated E or S0 
galaxies. This confirms the conclusion of Paper I that these
objects are spheroid-dominated systems, although with twice the sample of
objects and the better quality data we have here, it is now apparent that they
are not preferentially ellipticals but instead comprise a $\sim$35\%:65\% 
mix of E and S0s, respectively. Our profile fitting technique (\S 3.2) has
been crucial in this context, establishing that objects AC~114\#4, AC~114\#858, 
AC~114\#89 and AC~118\#4 are {\it bona fide} ellipticals rather than face-on
S0 galaxies.      

The purpose of Figure 6 has, up to this point, been to examine and discuss the
morphology of the `active' (SB, PSG, Sp, HDS) cluster members. The morphologies
of the 72 ``Psv'' and 8 ``Psv(pec)'' galaxies whose images have not been shown  
also deserve comment, in particular whether the former are consistent with 
their spectroscopic and photometric classification of being identical to  
present-day E or S0 galaxies (CS). By and large we find this to be the case 
with 59 (82\%) of the Psv types having Hubble types of E, E/S0, S0 or S0/a. 
The 13 remaining objects in this sample are all early-type (Sa, Sab, Sb) 
spirals; a representative collection of these objects is shown in Figure 6(d). 
Although it is only a small sample, it is interesting to note that all but one  
of the 8 Psv(pec) galaxies have an S0 or S0/a classification.  
 
Whilst we expect some fraction of the spiral population in clusters to be as 
red as the E and S0 populations (Butcher \& Oemler 1978b), such objects would 
be expected to have some on-going star formation and, strictly, should not be 
included in the ``Psv'' spectroscopic class due to their [OII] emission. 
On the other hand the strength of [OII] emission in present-day spirals, 
at least, is 
correlated with their color in the sense that the reddest objects have the 
weakest emission (Dressler \& Gunn 1982, CS). It could well be the case that 
we struggle to detect [OII] emission in such objects given the expected 
weakness of this feature (EW$\leq 5$\AA), the limited S/N of our spectra and
the limited size ($\ls 2$\,arcsec) of our spectroscopic aperture.  
A reexamination of the spectra shows this is likely in some but not all cases.  
Whereas objects CN\#253 and CN\#171 in Fig. 6(d) have spectra with a S/N 
insufficient to rule out [OII] emission at the EW$\leq 5$\AA\ level, objects  
CN\#509, CN\#537, CN\#232, and CN\#139 all have much superior spectra in which 
[OII] emission has to be absent at the EW$\leq 1$\AA\ level. Clearly high S/N 
spectroscopy of the light integrated over the full disk of these galaxies 
would be valuable in clarifying their emissive nature (or otherwise). 
Nonetheless, we note that all these galaxies have 
smoothly--textured and rather diffuse spiral arm structure similar to that  
of the `anaemic' gas--deficient spirals identified by van den Bergh (1976) in 
nearby rich clusters. This is suggestive of a family of `dead' spirals which
have possibly had their star formation extinguished through gas removal or gas  
starvation processes (Gunn \& Gott 1972; Larson, Tinsley \& Caldwell 1980). 

Finally, our systematic inspection of galaxy morphology according to 
star formation class has revealed numerous examples of dynamically interacting
systems amongst {\it all} types. We conclude this section by taking an
inventory of the numbers of M/I/T systems within the different spectroscopic  
classes and properly assessing the statistical significance of their rate of  
occurrence in each. The data are presented in Table 6 with the numbers for
each cluster listed individually and, to maximise the statistics, in combined 
form as well. The numbers in parentheses are those where the m, i, and t 
systems have also been included; for completeness, the numbers of objects 
recorded as having satellites (Sat classification) are also listed. 
Despite the numerous cases of major mergers/interactions found amongst
the SB and young PSG types (\S 5 and Fig. 6a), the galaxy samples involved
still remain small even when combined across all three clusters. When subjected 
to a formal error analysis (based on binomial small number statistics; Gehrels 
1986) we see that the higher incidence of dynamical interactions amongst
the SB and PSG classes in comparison to the other (Sp, HDS and Psv) types 
is yet to be established with any statistical significance. The most we can
conclude from our data is that dynamical interactions occur at a significantly
higher rate ($39^{+9}_{-8}$\%) across the blue active (SB+PSG+Sp) population 
in comparison to the Psv population (rate = $18\pm 3$\%). 
There is also a suggestion 
that the merger/interaction rate amongst the HDS population is at the same, 
lower level as the Psv population although better statistics are needed to  
confirm this. Table 6 also serves to highlight the fact that AC~103 is
somewhat of an exception in that its population of blue active galaxies 
contains no undisputed cases of mergers or interactions (M or I class objects). 
Whilst the statistics are even poorer in this individual case, it is of note  
that AC~103 has quite different global properties to AC~114 and AC~118, being  
significantly poorer (CN) and having a much lower lensing-inferred mass 
(Smail et al. 1997b; see \S 6). Perhaps even more relevant in this context 
is the virialisation process within AC~103 which, on the basis of its redshift  
distribution (see also \S 6), would appear to be more advanced than in the 
other two clusters with the consequence that merging/interacting systems are
possibly disrupted. 
 
\section{Discussion}

In this paper we have examined the morphological mix exhibited by the
galaxy populations in a sample of 3 distant clusters all at the one 
look--back time, $\tau\simeq 2.3\,h^{-1}$\,Gyr. We have studied
this both globally and in more detail by concentrating on
spectroscopically confirmed members categorised according to their
current or recent star formation activity.  Our analysis has provided a
detailed view of the morphology of galaxies within clusters at an
earlier epoch across a range of environments and cluster type. The challenge
remaining is to piece together the trends seen at this era across the
different classes of object and between the 3 clusters into a unified  
evolutionary picture which accounts for the observed changes in star formation  
activity and morphological content at subsequent times.  A
parallel aim in doing so is to glean clues about the physical
processes responsible for these changes.

A simple question to ask at the outset is whether the various sub-populations
identified in the distant clusters can be successfully described as
different stages of a single evolutionary cycle, rather than as a
collection of unrelated objects.  This issue has been addressed in
detail by B96 whose careful modelling showed that the SB, PSG and HDS
types, at least, are consistent with sequential phases of a single
evolutionary cycle, involving secondary star formation in a 
cross--section of cluster galaxies.  This conclusion was not
based solely on reproducing galaxy colors and spectra using
evolutionary burst models (as in CS), but also in accounting for the
{\it numbers} of objects observed in each phase, having properly taken
into account the evolutionary time-scales, the changes in luminosity
associated with the starburst, and the selection of the spectroscopic
samples. It was shown that the relative numbers of SB, PSG, HDS, and even
quiescent Sp types observed in AC~103, AC~118 and AC~114 were best
reproduced by a model population in which starbursts take place
stochastically in $\sim 30$\% of the objects over a $\sim 2$\,Gyr
period prior to the epoch of observation. Whilst this modelling places
only {\it mild} constraints on the detailed characteristics of the
starbursts, it demonstrated the viability of an evolutionary
connection between some subset of the SB, PSG, and HDS types. Galaxies
in these classes, therefore, need not be considered in isolation but
also as the possible antecedants/descendants of each other.

In the framework of this proposition our morphological study has
uncovered two striking trends which may be of major importance in
further testing the single cycle model of B96.  Firstly, the `older' 
($t_{\rm SF}>1$\,Gyr) PSG galaxies are all early--type {\it disk} (S0--Sb) 
systems whereas the HDS population comprises a mixture of these types
plus ellipticals as well.  
Secondly, the majority of the galaxies seen in the actual starburst 
phase or very soon after are involved in major interactions/mergers, thereby
suggesting that this process is the predominant trigger of this activity. 
The fact that we see numerous galaxies in the interacting,
pre-merger phase adds strong support to this. Not only do they provide
direct evidence of interactions/mergers taking place, but they also
imply the process to be widespread given the expected brevity ($\sim
0.5$\,Gyr; Mihos 1995) of this phase. However, an important question remains: 
can the observed morphologies of the evolved PSG and HDS types be reconciled  
with their formation through recent mergers and, if so, do we constrain the  
nature of these mergers?

Dealing first with the remnant morphologies expected from major
mergers, there are potentially two problems for this type of merger
picture: Firstly, there is a long--established result from simulations
that major mergers of disk galaxies have remnants which show pure de
Vaucouleurs ($r^{\frac{1}{4}}$) profiles, with little or no disk component  
remaining (Negroponte \& White 1983; Barnes 1988; Hernquist 1993). However, 
all the older PSG types are disk systems and E galaxies make up only 
$\sim 35$\% of the HDS population. This would indicate that the high 
proportion of galaxies with disks seen amongst the older PSG and HDS 
categories are not formed through major mergers. The second problem for the  
major merger scenario relates to the tidal debris that is predicted to
surround the remnants of major mergers for $\sim 1$\,Gyr after
coalescence (Mihos 1995). Mihos' simulations showed that although
faint, it should be clearly visible on WFPC-2 images of the depth
attained in this study, and yet we see no evidence of debris around the
younger PSG galaxies, where we would estimate that they are viewed
within the appropriate period, $t_{\rm SF}\leq 1$\,Gyr, after the
merger.   However, it is possible that the cluster environment has
additional effects on the visibility of this debris, perhaps stripping
it to form intracluster light, although such a process might also have
substantial effects on the galaxy in its own right.  Nevertheless,
unless there are shortcomings in our understanding of major mergers, it
seems unlikely that the majority of the PSG and HDS types are the
remnants of this kind of event. 

In terms of outcomes, ``minor'' mergers in which galaxies accrete satellite 
or dwarf galaxies can be quite different and less constraining. 
Although these events lack the visibility associated with the early stages
of a major merger, this may be more than compensated for in the much
larger pool of dwarf galaxies available to fuel minor mergers.  Moreover,
whilst minor mergers can significantly disturb the structure of a
galaxy, they can do so without destroying their disks (Quinn et al. 1993). 
Schweizer (1996) has even argued that they can contribute
to S0 formation. Although none of our galaxies show unambiguous
evidence of undergoing a minor merger, it is conceivable that features
such as the high central surface brightness of the SB galaxy AC114\#510
and the disturbed appearance of the PSG galaxy AC103\#132 (as well as a
number of the Sp types) result from this process, taking Quinn et al.'s
simulations as a comparative guide.

Another important consideration in this context is the selection effects 
that come into play in defining
our samples of SB, PSG and HDS objects. Of particular importance is the
sizeable increase in luminosity that accompanies starburst activity thereby
considerably enhancing the visibility of these objects in the SB and early 
PSG phases, but in contrast making them significantly fainter in the later
PSG and HDS phases (B96). We have attempted to minimise the bias this might
have on our spectroscopic samples by largely basing their selection on infrared 
$K'$ magnitudes (see \S 2.2). However, even in this band a $\sim 1$\,mag 
rise and fall in luminosity is seen at the beginning and end of the starburst  
phase (Fig. 2, B96); the galaxies in the top two rows of Fig. 6(a) would 
therefore seem destined to fade by this amount by the time they get to the 
late PSG phase. This would be sufficient for at least 4 of these galaxies  
(AC~114\#111, AC~103\#132, AC~118\#104, AC~114\#22 all with $17\leq K'\leq18$) 
to drop below the selection limit of our spectroscopic samples ($K'=18$). 
Hence if the infrared luminosities of the starburst systems we see here are  
representative of the population of cluster members that experience this  
phenomonen -- that is they are of {\it modest} luminosity with 
$L_{K'}\sim L_{K'}^{*}+1$\,mag -- then it is 
no surprise that the end--products are largely missing from our current 
samples, having faded to join the clusters' dwarf galaxy population. 

Despite these questions over the identification of the post--merger 
population, the fact remains that we see dynamical interactions actually 
taking place, not just amongst the starburst objects but across the whole  
cluster population (\S 5). This seems contrary to the notion of a 
``hot'' cluster environment where the relative velocities between galaxies 
exceed their internal values thus making strong interactions unlikely. However, 
this objection may well be a naive one for two reasons: Firstly, theoretical 
work by Mamon (1992) and Makino \& Hut (1997) has shown that  
interactions/mergers do still occur in rich clusters at levels sufficient to 
influence the evolution of galaxy systems despite the high relative velocities. 
Secondly, and perhaps more importantly, the assumption of a ``hot'' 
virialised environment in the 
clusters we observe here is not necessarily a valid one. If we look at the 
cluster redshift distributions assembled from all the available spectroscopic
data (Figure 7), all three show varying degrees of incompatibility with the
pure Gaussian form expected for a virialised population. AC~114 and AC~118,
in particular, show the greatest deviations in this context. Whilst the former
does contain a major gaussian component -- centred at $z=0.313$ and 
seemingly associated with the virialised core ($r\leq 150\,h^{-1}$\,kpc) 
region identified by lensing studies (Natarajan et al. 1997) -- a second 
velocity sub--component centred at $z=0.325$ is also clearly present. 
AC~118 shows no overall gaussian
behaviour whatsoever and, instead, appears to be a collection of at least
two velocity sub-structures. Indeed the unrelaxed state of AC~118 
is further emphasised by X-ray and gravitational lensing studies.  
Masses inferred from the weak lensing shear observed for background sources
in the field of AC~118 would, if the system was virialised, result
in a velocity dispersion $\sim$70\% smaller than that observed 
(Smail et al. 1997b). In addition, the shear map derived for AC~118 in this  
study together with the map of the hot X-ray gas obtained with the ROSAT PSPC  
(Bower et al. 1997) show this cluster to have two well resolved mass  
sub--components which are presumably coalescing.
 
As encouraging as this might be to understanding why the conditions within our clusters are favourable towards dynamical interactions, it still has to be
born in mind that the observed incidence of this phenomenon (20\% over our
3 clusters; Table 6) is considerably higher than that seen in any environment
nearby. Even in the Hickson compact groups -- recognised as being the most
favourable environments locally for galaxy interactions and mergers  -- show
a merger fraction of only 6\% (Zepf 1993). Clearly the evidence from this
study together with the similarly high fractions of interacting/merging
systems observed in other distant clusters (e.g. Paper I, Oemler et al. 1997) 
point to a significant evolution in the rate of interactions and mergers 
between these earlier epochs and the present day.

A useful theoretical framework in which such evolution might be explained and 
the assemblage of clusters through the merging of mass sub--components (as seen
in AC~118) is a major feature, is provided by the ``hierarchical'' clustering  
models described and developed in various ways by Bower (1991), 
Kauffmann (1995a,b) and Baugh et al. (1997). Of key importance in this 
scenario is that the distant clusters we observe here have had quite 
different evolutionary histories to their present--day counterparts, 
undergoing a much shorter and more turbulent period of merging prior to 
virialisation. As the above studies have shown, this has important evolutionary
implications for their galaxy populations, with the most massive systems
predicted to show the strongest effects. Evidence for this within our sample
is suggestive if not conclusive: The absence of any detectable 
lensing signal in the direction of AC~103 indicates that it has a central mass 
approximately an order of magnitude less massive than AC~118 and AC~114 
(Smail et al. 1997b, Natarajan et al. 1997). Notably, this cluster has the
highest core spiral fraction of our three clusters (see Table 5) and shows
the smallest incidence of interacting/merging amongst its blue population 
(see \S 5); however, its blue galaxy excess within $\sim 300\,h^{-1}$\,kpc of 
its center is no different to that of AC~118 and AC~114 (CS) and it has a 
similar population of active (SB, PSG, HDS) types. At the other extreme, 
studies of the `strong' and `weak' lensing effects seen towards AC~114 
(Smail et al. 1995a, Natarajan et al. 1997) indicate that it has a very massive 
and deep central ($r\ls 150\,h^{-1}$\,kpc) potential and it is in this very
region where we see the most evolved morphological mix within our 3 
clusters (\S 4.2). Clearly a bigger sample of clusters with well determined 
lensing--based mass distributions are needed to better clarify such trends.  

Finally, on the question of mechanisms, we are faced with a clear dichotomy 
amongst the population of active cluster members: there is a distinct subset
of objects undergoing major interactions/mergers but there is an equally
conspicuous collection of objects where this process is unlikely to have played 
any part in their recent evolution. For the former, the interacting/merging  
process provides a natural mechanism for triggering and then truncating a 
burst of star forming activity and, at the same time, producing systems with 
a spheroid morphology. For the latter -- particularly the population of normal 
disk systems (Fig. 6a,b) -- it would seem some other mechanism is responsible 
for the demise, at least, of their star formation activity. Numerous 
alternative mechanisms have been suggested and discussed in the literature: 
ram pressure stripping (Gunn \& Gott 1972, Kent 1981), galaxy infall and  
interactions with the intracluster medium (Bothun \& Dressler 1986), tidal  
compression by the cluster gravitational potential (Byrd \& Valtonen 1990, 
Valluri 1993), and ``galaxy harassment'' via high speed impulsive encounters 
(Moore et al. 1996).

The recent study by Abraham et al. (1996) of the rich cluster Abell~2390 at
$z=0.228$ has refocussed attention on galaxy infall, showing it to be a process 
of major importance in this cluster's development and the evolution of its 
galaxy population. Their evidence was based on comprehensive color, spectral, 
dynamical and coarse morphological information on galaxy members going out to very  
large radii ($r\sim 3$\,Mpc), which revealed gradients in these properties 
consistent with the cluster being fed and built up by an infalling population 
of field galaxies. If what is seen in A~2390 is a widespread phenomenon in
clusters, then the Abraham et al. study would indicate we should at least
see the remnants of this infalling population even within the much smaller  
fields studied here. Given the infallers should be predominantly normal 
spirals whose star formation is halted upon entry into the cluster, the 
remnants will be identifiable via their ``truncated spiral'' type 
spectra/colors and morphologies consistent with the later phases of a 
spiral~$\rightarrow$~anaemic spiral~$\rightarrow$~S0 transformation sequence. 
Certainly these characteristics have been recognised amongst our cluster
members. As discussed in the previous section, some of the older PSG types
shown in the bottom 3 rows of Fig. 6(a) have spectra and colors equally 
well explained by truncated star formation in a normal spiral. In addition 
they are all disk galaxies with either an S0 or early-type spiral morphology
as might be expected. Another set of candidates is the `dead' spirals 
with their passive--type spectra/colors and anaemic spiral morphology 
(Fig. 6d). Importantly though, we also see in our clusters spirals undergoing 
normal star formation which raises the question as to why they have managed to 
survive? Their distribution in redshift (Fig. 7) gives no (limited) indication  
that they are peripheral objects and hence perhaps early infallers. 
Alternatively, we might be seeing them after just one or two crossings 
of the cluster, which are insufficient to completely wipe out their star
formation activity.  

Turning to the galaxy harassment scenario (Moore et al. 1996), it is  
particularly interesting as it makes quite specific predictions as to the  
duration of the visible phase of this process and the morphological signatures  
that are expected to be seen. Specifically, Moore et al's simulations 
of a model spiral galaxy falling into a cluster show the harassment phase
to last for several billion years during which the galaxy's spiral arms are
drawn out of its disk and, as such, are subjected to severe  
tidal distortions (as seen in their Figure 3b). Given the timescale together
with the fact that this is expected to be a continuous process, a high 
fraction of blue disk galaxies showing this type of distortion should be
seen over a wide range in redshift. An inspection of our distant  
cluster galaxy morphologies reveals only 3 possible cases of spirals in this  
harassed state (AC~118\#438, AC~114\#1.147, and AC~103\#62), but these examples 
are far from dramatic. Furthermore, according to Moore et al., disturbed 
spirals are likely to be accompanied by starbursts, but our SB category does 
not show any distorted spirals. While this casts some doubt as to the 
widespread operation of galaxy harassment within our clusters, it is perhaps  
premature to rule it out entirely. The main reason for this is that Moore et 
al's simulations have explored only a small region of parameter space in terms 
of the properties (luminosity, morphology, etc) of the harassed galaxy; 
more comprehensive simulations of this process will better establish its
importance.  

The data presented here on three clusters of {\it intermediate} redshift
further emphasise the {\it rapid} evolution in morphological content and
star formation activity that has occurred in rich clusters over the last 
one third of a Hubble time. Fundamentally, however, this tracking of cluster
galaxy morphology back in time both by our study and even further by the
MORPHS study (Smail et al. 1997a, Dressler et al. 1997a) has revealed the
changes to be one of morphological mix rather than one of basic galaxy
structure: the same basic Hubble types we see today are in place in clusters
at $z=$0.3--0.5. Nonetheless, the epoch when this basic structure formed
in cluster galaxies would appear to be within grasp. Observations of clusters 
at $z\sim 1$ with $HST$ by Dickinson (1997) and Faber et al. (1997) reveal
galaxy populations dominated by chaotic and fragmented systems bearing no 
resemblance to the normal Hubble types seen in the clusters at $z<0.5$.
Clearly the interval $0.6\leq z\leq 1$ merits similarly detailed attention if
we are to fully unravel the morphological development of galaxies within the
rich cluster enviroment.

\section{Summary}

This paper has presented deep, $HST$--based high resolution imagery of 3 rich  
but diverse clusters at $z=0.31$ from which detailed morphologies of their  
galaxy populations have been derived and analysed with reference to  
ground--based spectroscopy and optical--infrared multi--band photometry. 
The primary results of our study can be summarised as follows:

\noindent
(1)\,Our 3 clusters, each of which exhibit a significant blue 
BO excess, provide further evidence that this effect is
linked to the presence of greater numbers of spiral galaxies
in these environments at earlier epochs. We find these spirals to be generally
of normal Hubble type spanning the full Sa--Sdm/Irr range and, collectively, are
present in numbers up to $\sim$4 times higher than that seen in the same 
high density environments nearby. The galaxies directly responsible for the 
blue BO excess are predominantly of mid to late (Sb--Sdm) Hubble type. 

\noindent
(2)\,Some indications as to the dependency of this spiral galaxy excess 
on global cluster environment are provided by our study. In the core 
($r\ls 200\,h^{-1}$\,kpc) regions we find the highest spiral fraction (35\%) 
to be in the least massive, most irregular cluster in our sample (AC~103). 
In contrast, the morphological mix closest to that seen at the same densities  
nearby is found in the core of our most massive and regular cluster, AC~114.  
Furthermore, our study of the outer ($r\gs 200\,h^{-1}$\,kpc) regions of this 
latter cluster, where the projected galaxy density has dropped by a factor 
of $\sim 3$ relative to the core, reveals spirals to be in excess by a 
similar factor. 

\noindent     
(3)\,The higher fractions of spiral galaxies in these clusters is accompanied
by a deficit in either or both E and S0 galaxies. Although still substantial, 
the deficit in S0 galaxies seen here at $z=0.31$ is significantly lower than 
that seen in rich clusters at $z\sim 0.5$ (Smail et al. 1997a, Dressler et al. 
1997a), indicating a steep downward trend consistent with a quite rapid 
production of S0s in cluster cores since $z\sim 0.5$. 

\noindent
(4)\,On average, one in every five cluster members (both blue and red) show
morphological signatures indicative of dynamical interactions. A similar  
fraction of the spirals in our clusters also show structural abnormalities 
in which their arms appear to be mildly asymmetric or distorted. Examination
of the spatial distribution of both sets of objects within our clusters shows 
no convincing evidence that they are clustered in any way or avoid
particular regions of the cluster. The fraction of galaxies involved in dynamical interactions appears to remain constant with radius from the 
cluster center.

\noindent
(5)\,Our extensive spectroscopy and multi--band photometry of the clusters
in our sample has allowed us to focus on the morphology of those cluster
members identified as being currently active in star formation or having been 
so in their recent past. The very blue galaxies either undergoing a major  
starburst or having just recently completed one are predominantly systems 
involved in {\it major} mergers. Dynamical interactions appear to be a common 
cause of this most extreme form of star formation activity seen in these   
clusters. With the caveat that our statistics are small, the galaxies involved  
are mostly of modest luminosity ($L_{K'}\sim L_{K'}^{*} + 1$\,mag) even in  
this brightened phase; in their later faded state they appear destined to 
become dwarfs, too faint to be included in magnitude--limited spectroscopic  
samples such as ours. 
The galaxies with ongoing star formation occuring at the rates
typical of normal nearby spirals, are mostly normal Sb--Sdm/Irr Hubble 
types. For the later types in this group, their visibility on our WFPC-2 
images is due mainly to their compact, knotty regions of star 
formation; it is clear they would become quite diffuse, low surface brightness  
systems if their star formation were to cease. 

\noindent
(6)\,The galaxies seen at times $\gs 1$\,Gyr after their last 
major episode of star formation -- the blue ``PSG'' and red ``HDS'' objects --  
also show clear morphological trends. The former, whose spectra and colors
are consistent with them having either a starburst or ``truncated''  
spiral origin, are all normal early--type (S0--Sb) {\it disk} galaxies -- 
a morphology seemingly inconsistent with them having experienced a major 
merger, at least in their recent past. It would seem that they are mostly  
descendants of the active spiral population who have had their star formation 
curtailed by some other mechanism, thereby (as we see directly) feeding the  
early--type disk populations in these clusters. The HDS objects, interpreted 
from their spectra and colors as being the remnants of secondary star formation 
in old dormant systems, have morphologies consistent with this picture, being  
mostly normal, isolated E or S0 galaxies. 

\noindent
(7)\,Collectively, these results provide further evidence that the evolution 
in star formation activity and morphological content associated with the  
BO--effect involves not one but several physical processes. Only one of 
these -- galaxy-galaxy interactions and merging -- can we see at work 
directly, providing a natural mechanism for triggering and then truncating 
star formation and transforming galaxies morphologically into spheroidal 
systems. However, an almost equally conspicuous feature of our samples is 
the numbers of objects where this process could {\it not} have played any 
major part in altering the course of their star formation activity. Pinpointing  what processes are responsible in these cases is a more difficult challenge. 
To this end our study does provide some important clues: The normal disk 
morphology of the old ``PSG'' population together with the anaemic looking  
spirals found amongst the ``Psv'' types point to processes which bring 
star formation to a halt but leave the basic disk structure intact and 
largely unperturbed. This seems more consistent with the operation of 
mechanisms which affect just the gas supply (eg. ram-pressure stripping, 
galaxy infall) rather than all mass components of the galaxy (eg. tidal 
processes). Indeed as far as ``galaxy harassment'' is concerned (a prominent
example of the latter), we find no evidence of the population of severely  distorted spirals that is predicted if the operation of this mechanism is
widespread.

\acknowledgements

We wish to thank Ray Lucas for his expert support and guidance 
in gathering the data for this program and Jean-Paul Kneib for 
the use of the imaging data on AC~114. We thank our colleagues Richard 
Bower, Alan Dressler, Gus Oemler, Bianca Poggianti and Alfonso Arag\'{o}n-Salamanca for many stimulating discussions throughout the
course of this work. WJC acknowledges the financial support of the Australian 
Research Council, the Australian Department of Industry, Science \& Tourism, 
and SUN Microsystems.

\clearpage

\begin{deluxetable}{lccccccc}
\tablenum{1}
\tablewidth{0pt}
\tablecaption{Log of $HST$ Observations \label{tab1}}
\tablehead{
\colhead{}&\colhead{}&\colhead{}&\colhead{}&\colhead{}&\colhead{}&
\colhead{}&\colhead{Exposure}\\
\colhead{Cluster}&\colhead{}&\colhead{R.A.(1950)}&\colhead{Decl.(1950)}&
\colhead{Date}&\colhead{Filter}&\colhead{P.A.(V3)}&\colhead{(s)} }
\startdata
AC~118&&00 11 47.9&-30 39 58.6&1994 Nov 22&F702W&217.0&6500 \nl
AC~103&&20 52 51.5&-64 50 26.3&1995 Jun 10&F702W&135.0&6500\nl
AC~114&\#1&22 56 02.0&-35 05 05.3&1996 Jan 05&F702W&270.0&16800\nl
      &\#2&22 56 02.0&-35 03 16.3&1996 Jan 05&F702W&270.0&16800\nl
      &\#3&22 56 10.6&-35 05 05.8&1996 Jan 06&F702W&270.0&16800\nl
      &\#4&22 55 53.3&-35 03 16.0&1996 Jan 07&F702W&270.0&16800\nl
\enddata
\end{deluxetable}

\clearpage

\begin{deluxetable}{lccccc}
\tablenum{2}
\tablewidth{0pt}
\tablecaption{Log of Spectroscopic Observations \label{tab2}}
\tablehead{
\colhead{}&\colhead{}&\colhead{}&\colhead{}&\colhead{Exposure}&
\colhead{Seeing}\\
\colhead{Cluster}&\colhead{Telescope/Instr}&\colhead{Mask}&
\colhead{Date}&\colhead{(s)}&\colhead{(arcsec)}  }
\startdata
AC~103&AAT/LDSS-1&\#1&1995 Aug 28&14000&1.5 \nl
      &AAT/LDSS-1&\#2&1995 Aug 29&10000&1.7 \nl
AC~118&AAT/LDSS-1&\#1&1995 Aug 29&10000&1.3--2.0 \nl
      &AAT/LDSS-1&\#2&1995 Aug 28&15000&1.4 \nl
AC~114&AAT/LDSS-1&\#1&1995 Aug 29&10000&1.4 \nl
      &NTT/EMMI&`bright'&1995 Oct 22&7200&0.8 \nl
      &NTT/EMMI&`faint'&1995 Oct 23,25&7200&0.8--1.6 \nl
\enddata
\end{deluxetable}

\clearpage

\begin{deluxetable}{lccccccccc}
\tablenum{5}
\tablewidth{0pt}
\tablecaption{Morphological Content of Clusters \label{tab5}}
\tablehead{
\colhead{Cluster}&\colhead{Region}&\colhead{N$_{\rm cl}$}
&\multicolumn{3}{c}{Observed\tablenotemark{a}}&\colhead{}&\multicolumn{3}{c}{Predicted}\\
\colhead{}&\colhead{}&\colhead{}&\colhead{\%E}&\colhead{\%S0}&
\colhead{\%Sp}&\colhead{}&\colhead{\%E}&\colhead{\%S0}&
\colhead{\%Sp}  }
\startdata
AC~103&WFPC-2&84&34$^{+2}_{-3}$&31$^{+3}_{-2}$&35$^{+2}_{-3}$&&42&49&9\nl
\nl
AC~118&WFPC-2&100&27$^{+3}_{-2}$&51$^{+2}_{-2}$&22$^{+2}_{-3}$&&45&50&5\nl
\nl
AC~114&$0'\leq r<0.5'$&24&55$^{+8}_{-8}$&39$^{+8}_{-8}$&6$^{+15}_{-6}$&&49&49&2\nl
&$0.5'\leq r<1.0'$&36
&38$^{+6}_{-5}$&40$^{+6}_{-5}$&22$^{+7}_{-4}$&&36&47&17\nl
&$1.0'\leq r<1.5'$&64
&19$^{+3}_{-3}$&29$^{+3}_{-3}$&52$^{+3}_{-4}$&&31&46&23\nl
&$1.5'\leq r<2.0'$&71
&12$^{+4}_{-2}$&24$^{+3}_{-2}$&64$^{+3}_{-3}$&&30&47&23\nl
&$2.0'\leq r<2.5'$&46
&26$^{+5}_{-4}$&29$^{+5}_{-3}$&45$^{+4}_{-5}$&&32&46&22\nl
&$2.5'\leq r<3.0'$&21
&14$^{+13}_{-6}$&24$^{+11}_{-8}$&62$^{+8}_{-10}$&&31&46&23\nl
&WFPC-2\tablenotemark{b}&117&44$^{+1}_{-2}$&37$^{+2}_{-2}$&19$^{+2}_{-2}$&&42&48&10\nl
\tablenotetext{a}{Errors are 1$\sigma$ confidence limits calculated using the 
tables published by Gehrels(1986)}
\tablenotetext{b}{WFPC-2 sized field centred on the cluster}
\enddata
\end{deluxetable}

\clearpage

\begin{deluxetable}{lcccccccc} 
\tablenum{6}
\tablewidth{0pt}
\tablecaption{Incidence of Dynamical Interactions \label{tab6}}
\tablehead{
\colhead{}
&\multicolumn{5}{c}{Numbers of objects}&\colhead{}
&\multicolumn{2}{c}{Percentages\tablenotemark{a}}\\ 
\colhead{Spectral Class}&\colhead{Total}&\colhead{M}&
\colhead{I}&\colhead{T}&\colhead{Sat}&\colhead{}&\colhead{(M+I+T)}
&\colhead{Sat}  }
\startdata
\multicolumn{9}{c}{AC~103}\nl
Psv&17&0(0)&0(0)&1(3)&4&&6(18)&24\nl
HDS&4&0(0)&0(0)&1(2)&1&&25(50)&25\nl
PSG&2&0(1)&0(0)&0(0)&1&&0(50)&50\nl
SB&0&0(0)&0(0)&0(0)&0&&0(0)&0\nl
Sp&3&0(1)&0(0)&1(2)&0&&33(100)&0\nl
\nl
\multicolumn{9}{c}{AC~118}\nl 
Psv&21&0(0)&5(5)&2(2)&3&&33(33)&14\nl
HDS&3&0(0)&0(0)&0(0)&1&&0(0)&33\nl
PSG&3&1(1)&1(1)&0(0)&0&&67(67)&0\nl
SB&0&0(0)&0(0)&0(0)&0&&0(0)&0\nl
Sp&2&0(1)&0(0)&0(0)&0&&0(50)&0\nl
\nl
\multicolumn{9}{c}{AC~114} \nl
Psv&42&1(1)&4(4)&1(2)&5&&14(17)&12\nl
HDS&7&0(0)&1(1)&0(0)&1&&14(14)&14\nl
PSG&7&1(1)&1(1)&0(0)&0&&29(29)&0\nl
SB&3&0(0)&2(2)&0(0)&0&&67(67)&0\nl
Sp&3&1(1)&1(1)&0(0)&0&&67(67)&0\nl
\tablebreak
\multicolumn{9}{c}{Overall}\nl 
Psv&80&1(1)&9(9)&4(7)&12&&$18^{+3}_{-3}$($21^{+3}_{-2}$)&$15^{+3}_{-2}$\nl
HDS&14&0(0)&1(1)&1(2)&3&&$14^{+19}_{-9}$($21^{+18}_{-10}$)&$21^{+18}_{-10}$\nl
PSG&12&2(3)&2(2)&0(0)&1&&$33^{+18}_{-13}$($42^{+16}_{-15}$)&$8^{+26}_{-7}$\nl
SB&3&0(0)&2(2)&0(0)&0&&$67^{+27}_{-42}$($67^{+27}_{-42}$)&$0^{+46}$\nl
Sp&8&1(3)&1(1)&1(2)&0&&$38^{+23}_{-20}$($75^{+16}_{-24}$)&$0^{+21}$\nl
PSG+SB+Sp&23&3(6)&5(5)&1(2)&1&&$39^{+9}_{-8}$($57^{+8}_{-9}$)&$4^{+16}_{-4}$\nl
\nl

Total&117&4(7)&15(15)&6(11)&16&&$21^{+2}_{-1}$($28^{+2}_{-1}$)&$14^{+2}_{-2}$\nl
\tablenotetext{a}{Errors, where quoted, are 1$\sigma$ confidence limits  
calculated using the tables published by Gehrels(1986)}
\enddata
\end{deluxetable}

\newpage

\begin{figure}[p]
\centerline{{\psfig{file=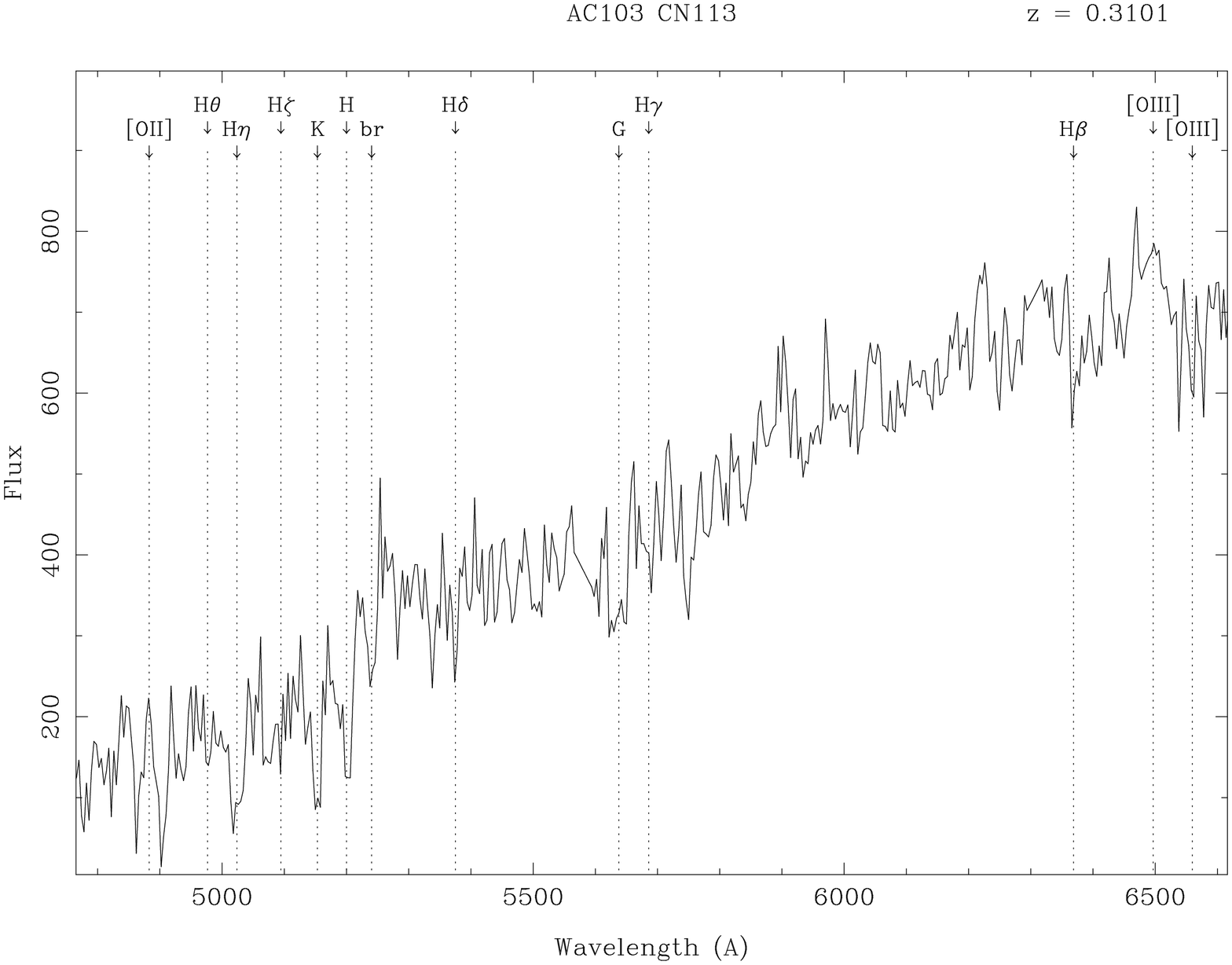,height=80mm,width=110mm,angle=270}}}

\centerline{{\psfig{file=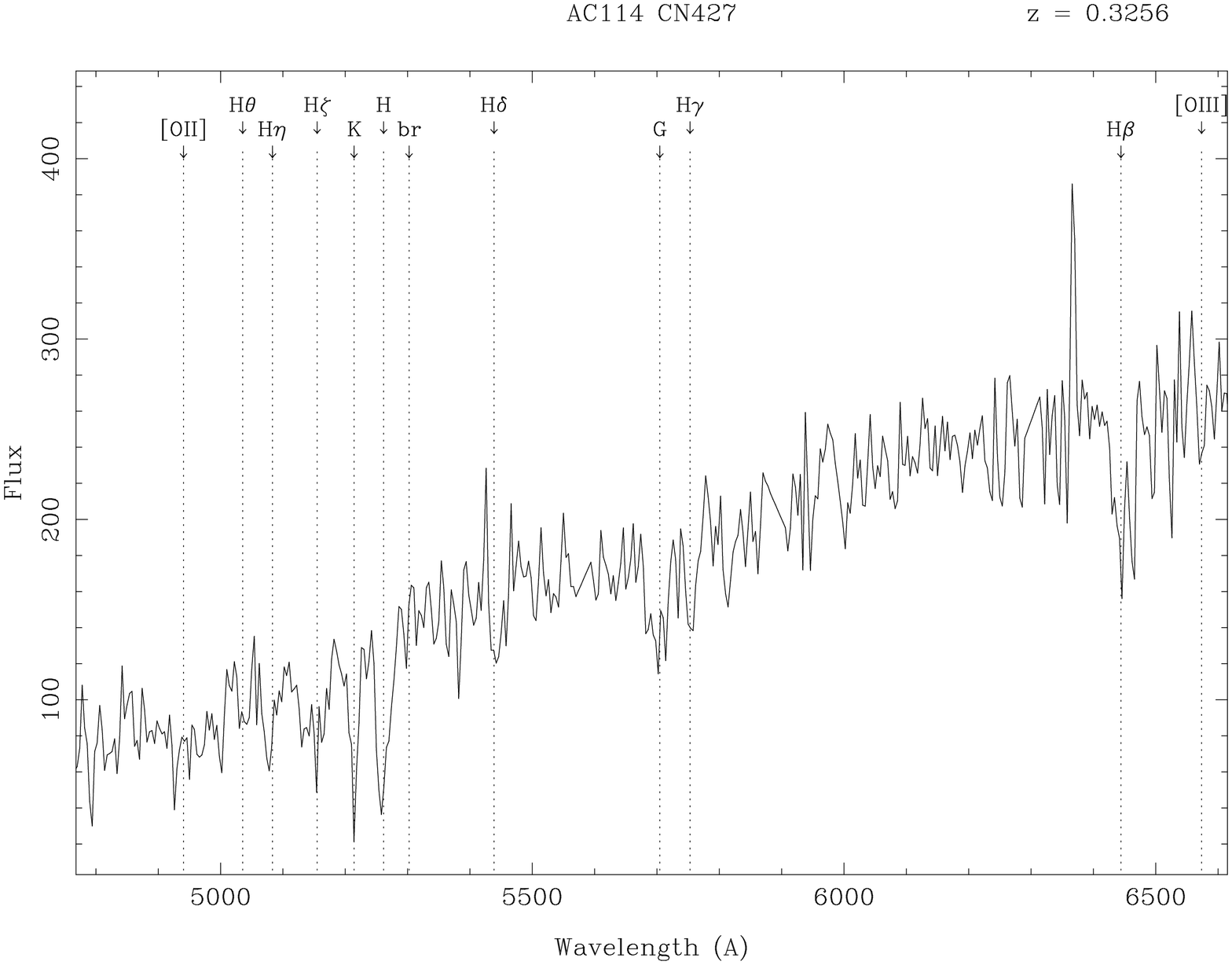,height=80mm,width=110mm,angle=270}}}

\centerline{{\psfig{file=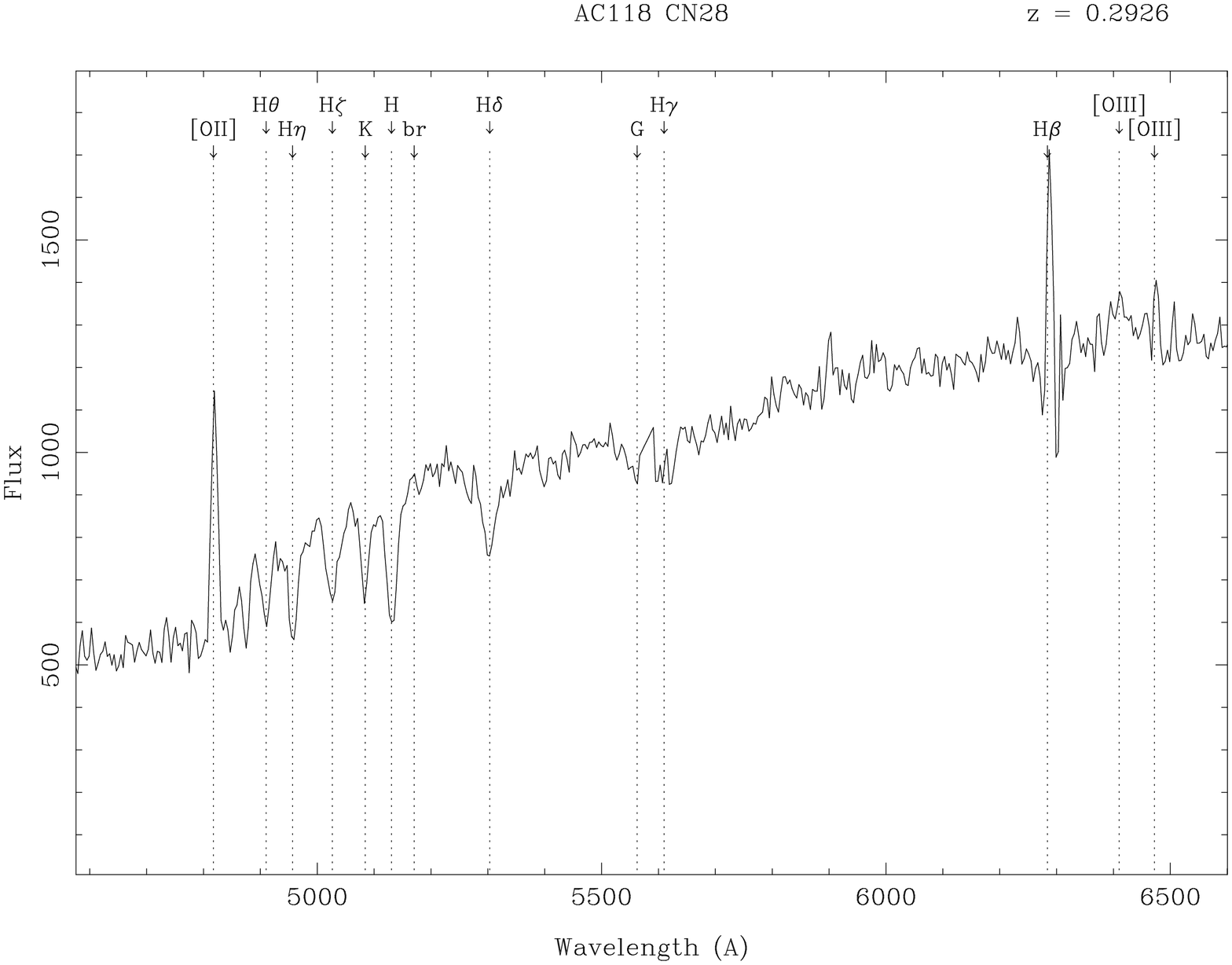,height=80mm,width=110mm,angle=270}}}
\caption{Representative examples of new spectra obtained for members within 
our 3 clusters at $z=0.31$.}
\end{figure}

\begin{figure}[p]
\centerline{{\psfig{file=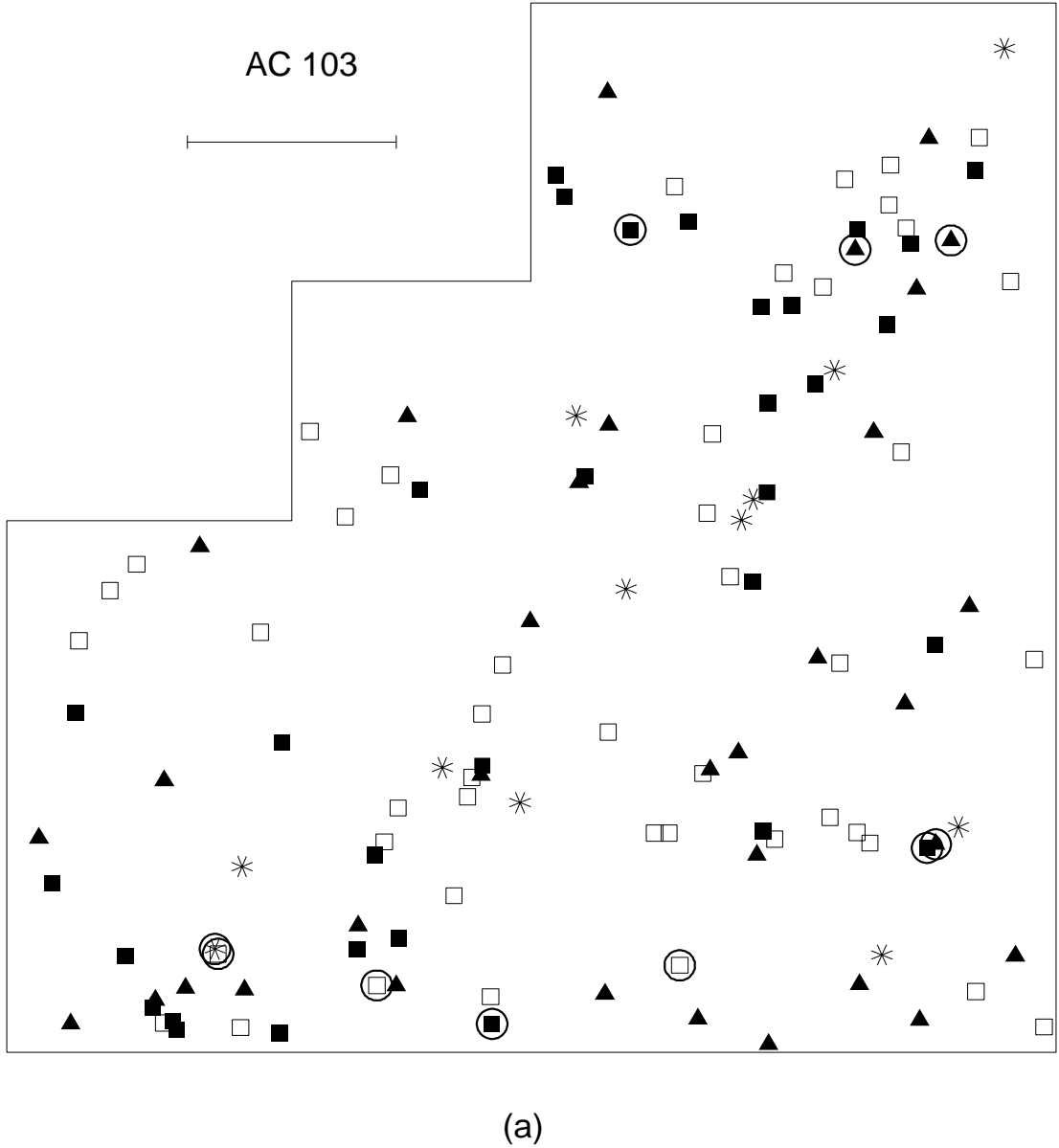,height=201mm,width=204mm}}}
\caption{Plots showing the position and morphology of the galaxies classified
in our WFPC-2 fields: (a)\,AC~103, (b)\,AC~118, (c)\,AC~114. The different 
morphological types are represented as follows: E -- {\it filled squares}, 
S0 -- {\it filled triangles}, Sp -- {\it open squares}, Pec/Comp -- {\it 
asterisks}, suspected lensed images -- {\it stars}; objects involved in 
dynamical interactions (M, I or T classification) are {\it circled}. The scale bar is 0.5\,arcmin in length.}
\end{figure}
\begin{figure}[p]
\centerline{{\psfig{file=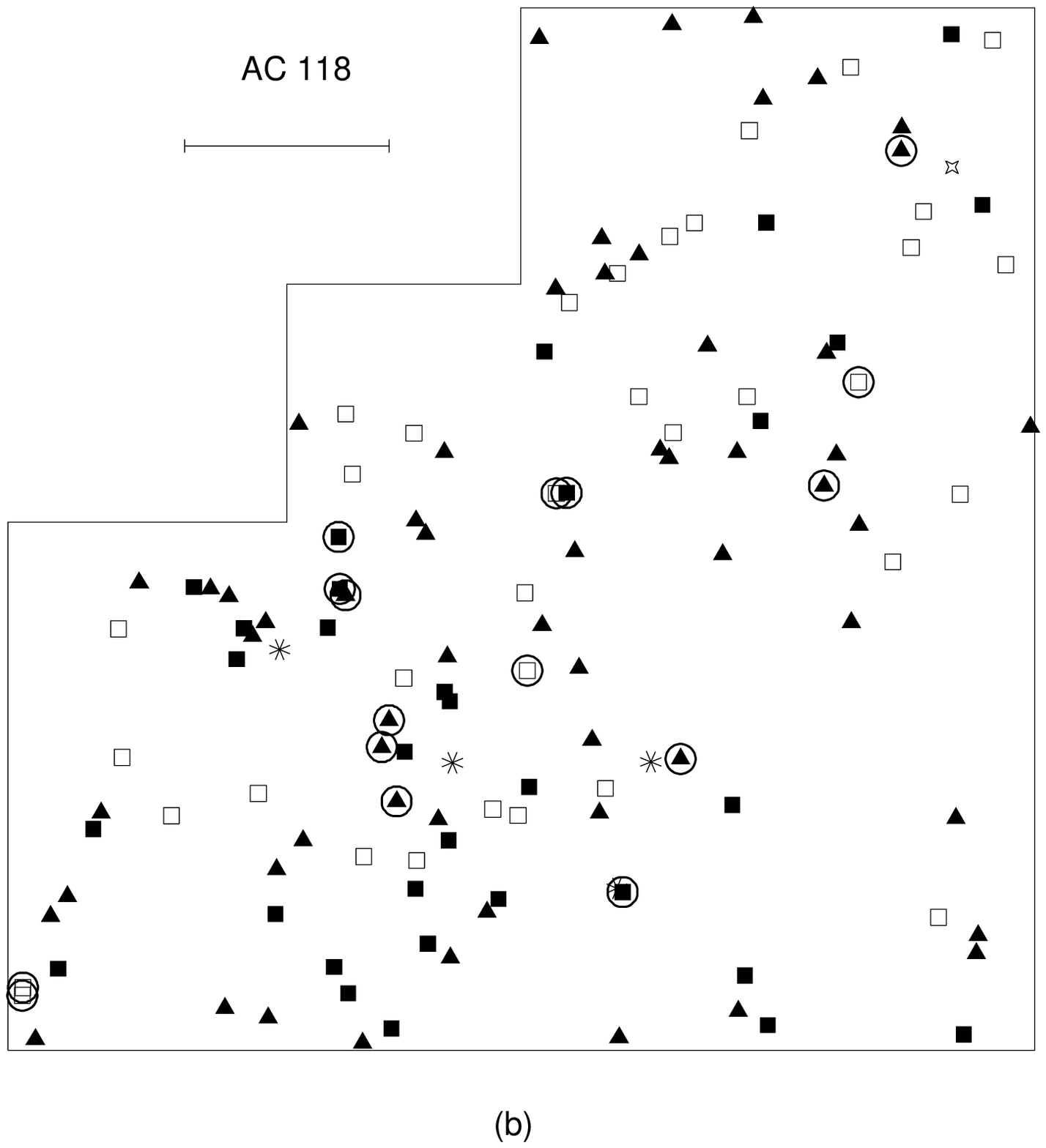,height=201mm,width=204mm}}}
\end{figure}
\begin{figure}[p]
\centerline{{\psfig{file=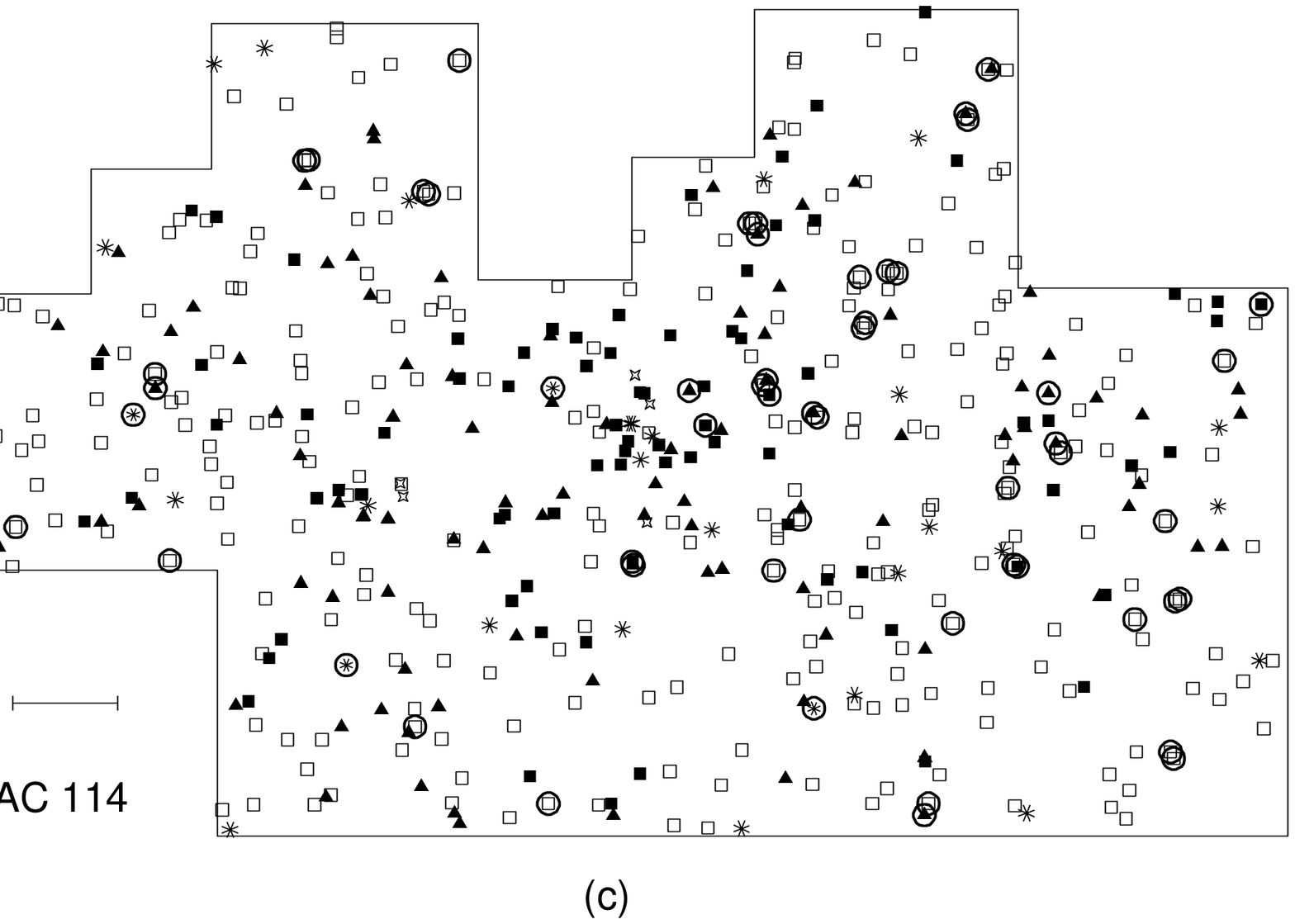,angle=90,height=204mm,width=201mm}}}
\end{figure}

\begin{figure}[p]
\centerline{{\psfig{file=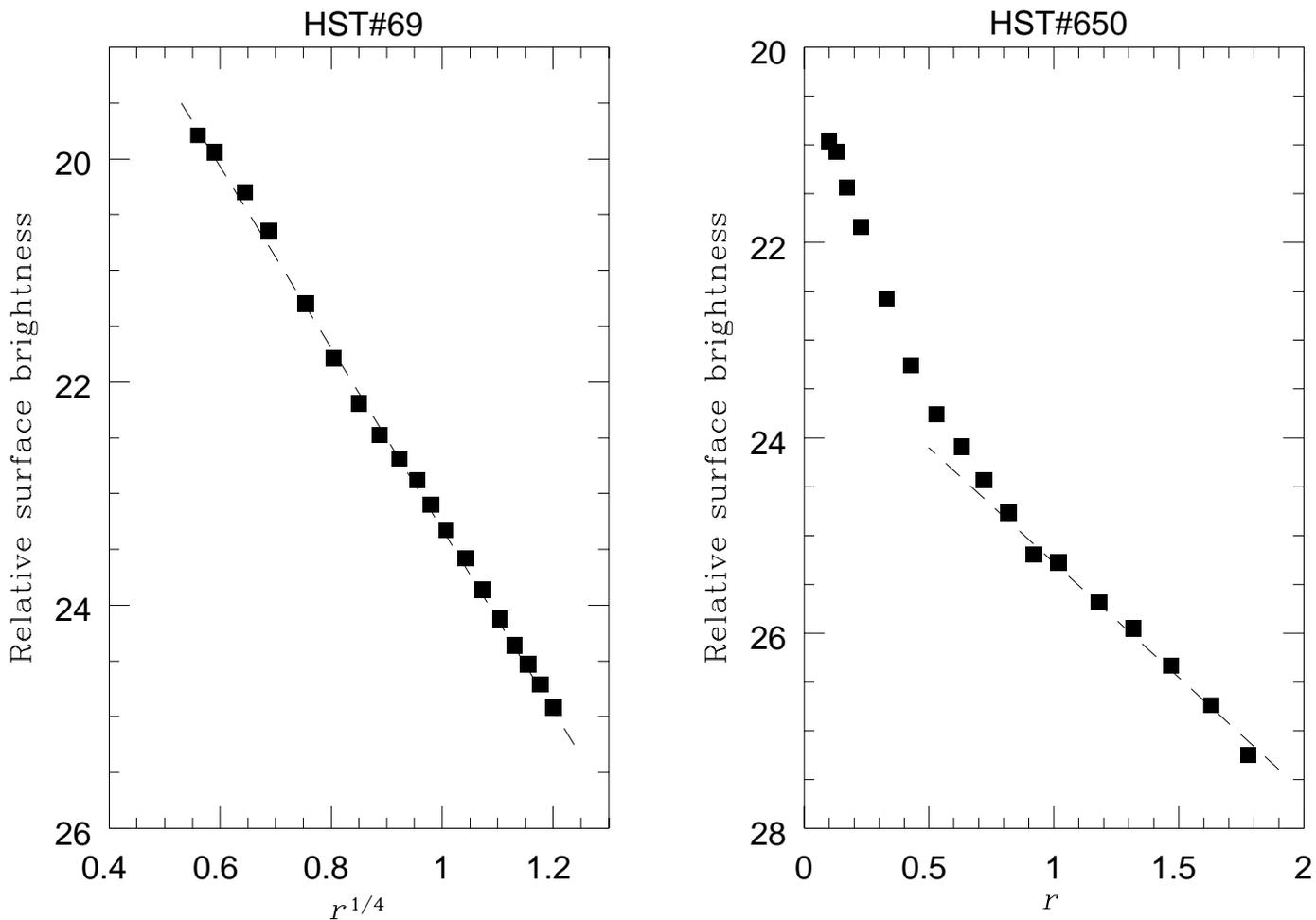,height=200mm,width=200mm}}}
\vspace{-1in}
\caption{Example surface brightness profiles for two galaxies in AC~103. 
{\it Left-hand panel:} relative surface brightness (in mags\,arcsec$^{-2}$) 
versus $r^{\frac{1}{4}}$ (in arcsec) for HST\#69 with a pure de Vaucouleur  ($r^{1/4}$) profile shown by the {\it dashed line}; {\it Right-hand panel:} relative surface brightness versus $r$ (in arcsec) for HST\#650 indicating an S0 classification on the basis of its exponential disk profile at large radii ({\it dashed line}) (see text for details).}
\end{figure}

\begin{figure}[p]
\centerline{{\psfig{file=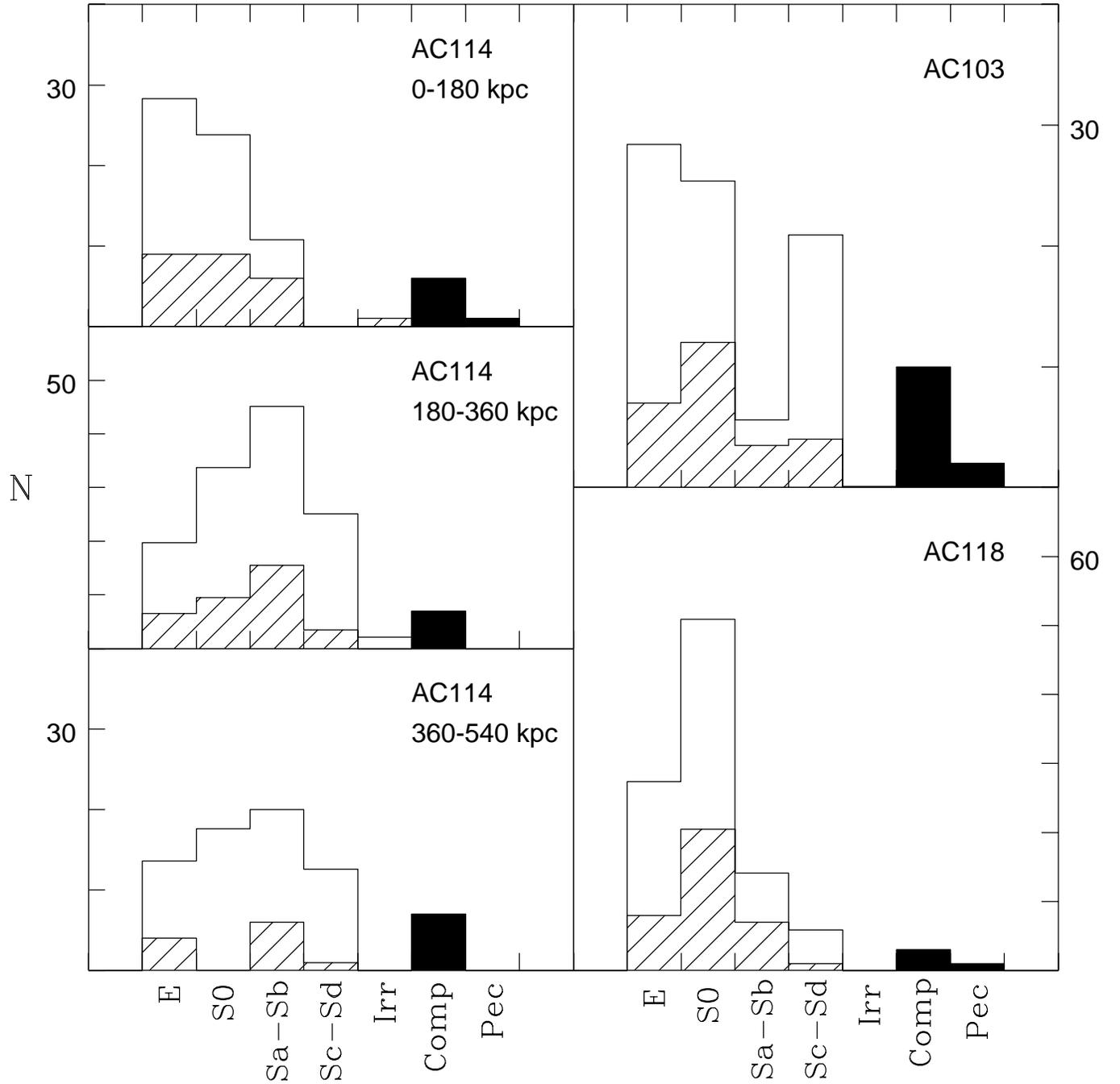,height=190mm,width=180mm}}}
\caption{The distribution of galaxies classified within our WFPC-2 fields according to Revised Hubble type. {\it Open} histograms represent our complete magnitude--limited samples corrected for field galaxy contamination; {\it  cross--hatched} histograms represent spectroscopically confirmed cluster members. The numbers of ``Comp'' and ``Pec'' types are also shown ({\it solid histograms}).}
\end{figure}

\begin{figure}[p]
\centerline{{\psfig{file=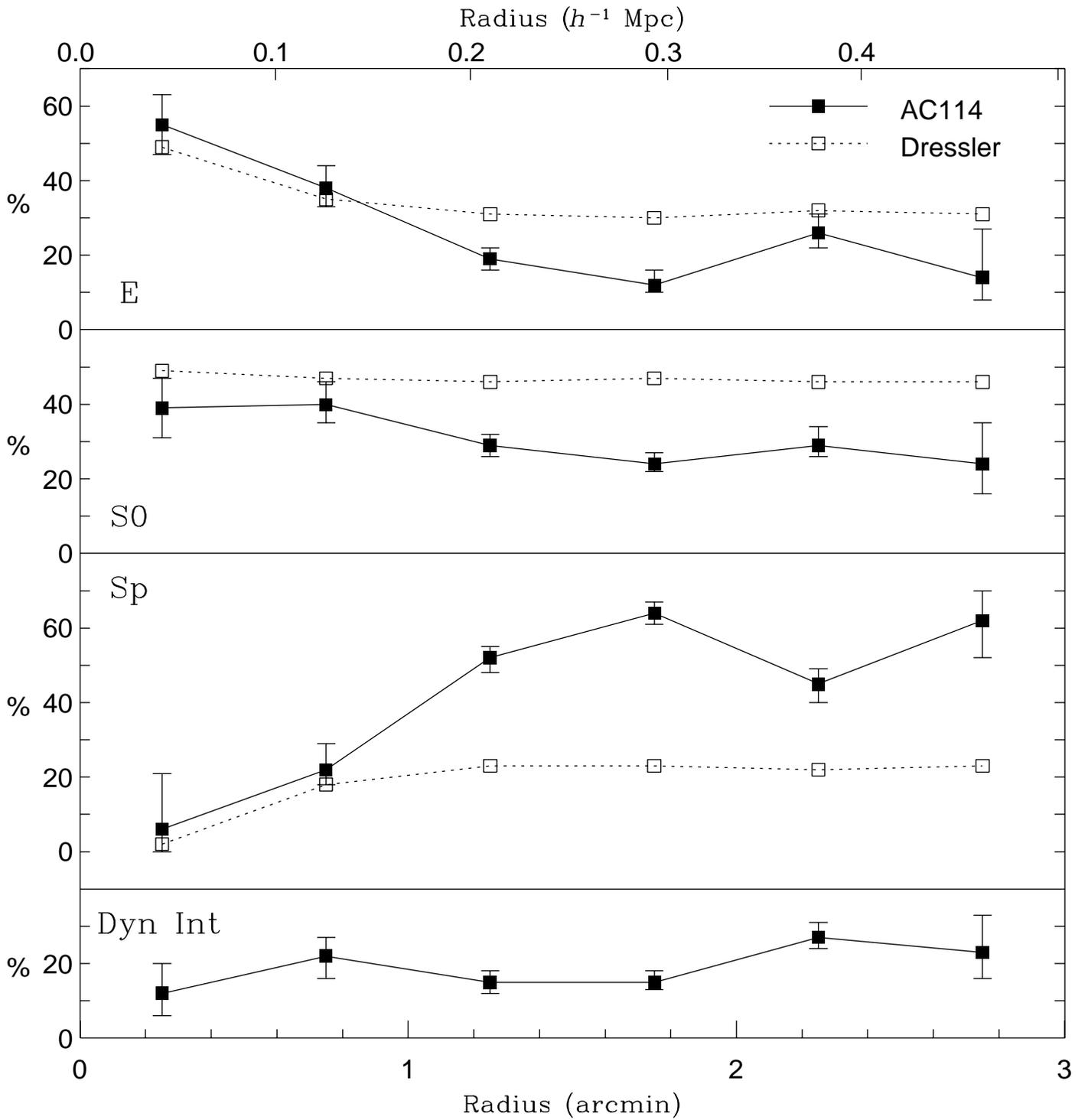,height=201mm,width=204mm}}}
\caption{The fraction of different morphological types as a function of radius in the cluster AC~114.}
\end{figure}

\begin{figure}[p]
\centerline{{\psfig{file=cess2-fig6a.eps,height=270mm,width=230mm}}}
\vspace{-1in}
\caption{WFPC-2 images of cluster members in the different star formation categories. The tick marks around the periphery of each image are at 
1\,arcsec intervals. (a)\,``SB'' ({\it top row}) and ``PSG'' types.}
\end{figure}

\setcounter{figure}{5}

\begin{figure}[p]
\centerline{{\psfig{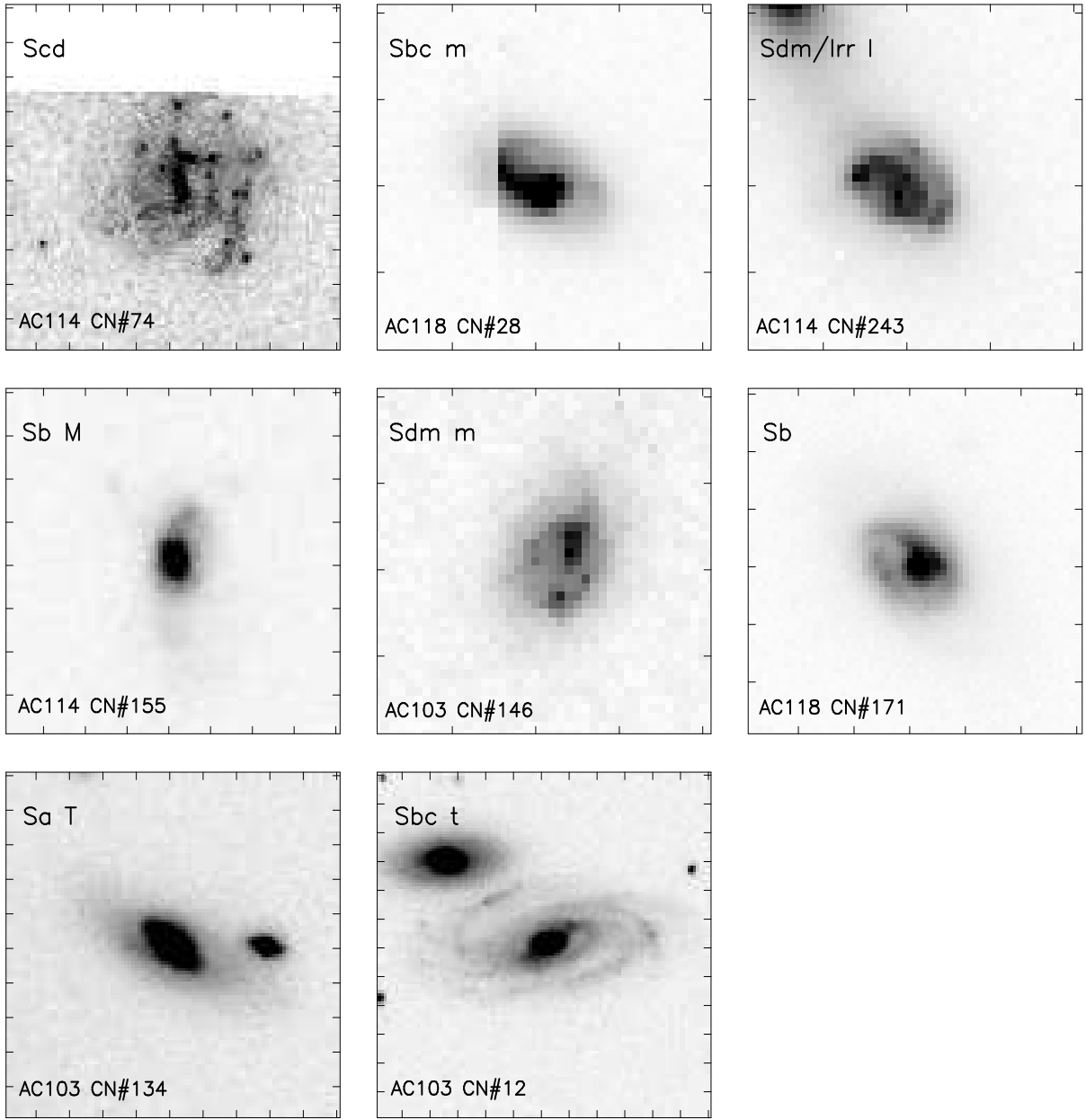}}}
\vspace{-2.5in}
\caption{(b)\,``Sp'' types.}
\end{figure}

\setcounter{figure}{5}

\begin{figure}[p]
\centerline{{\psfig{file=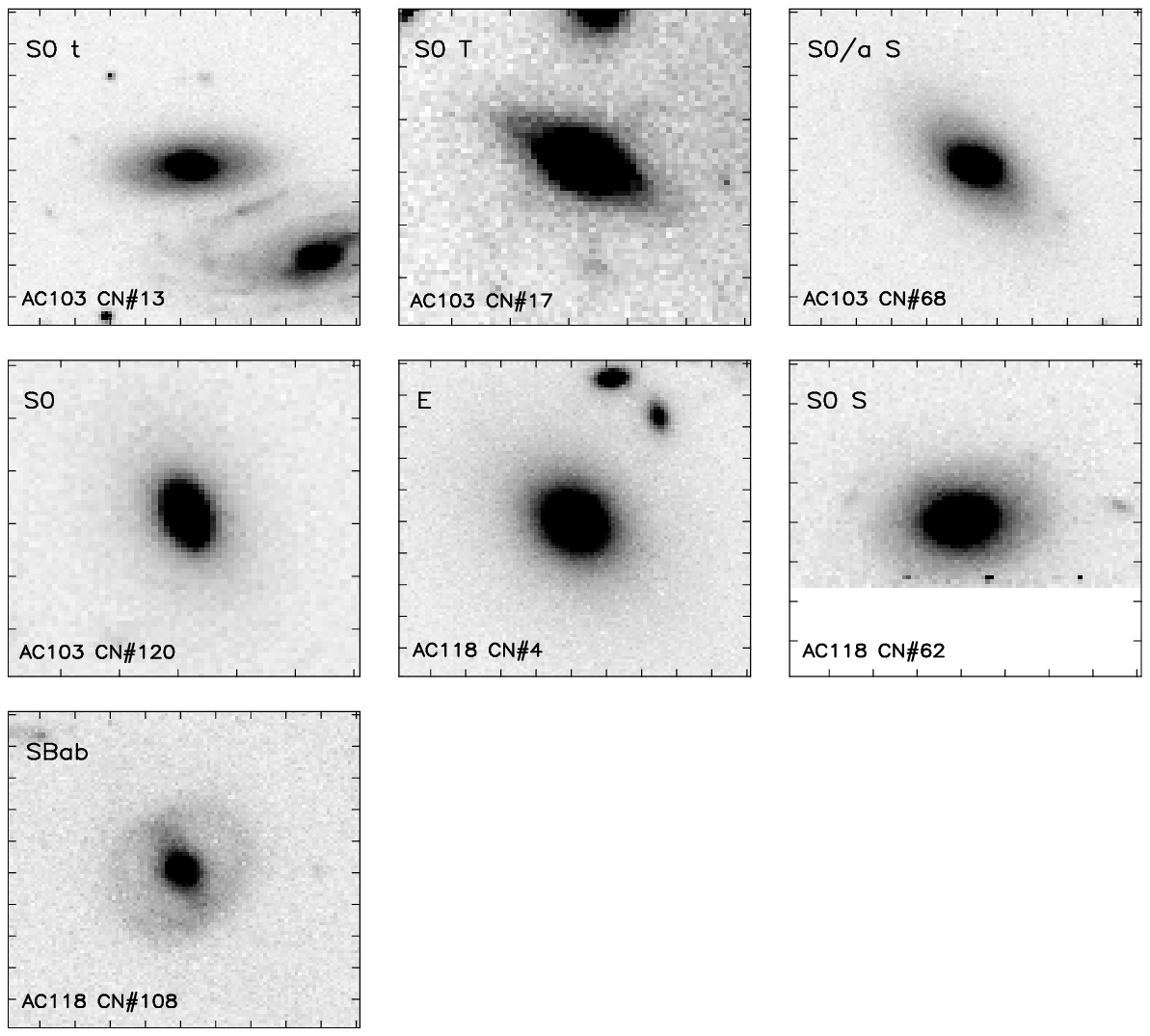,height=240mm,width=204mm}}}
\caption{(c)\,``HDS'' types.}
\end{figure}

\setcounter{figure}{5}

\begin{figure}[p]
\centerline{{\psfig{file=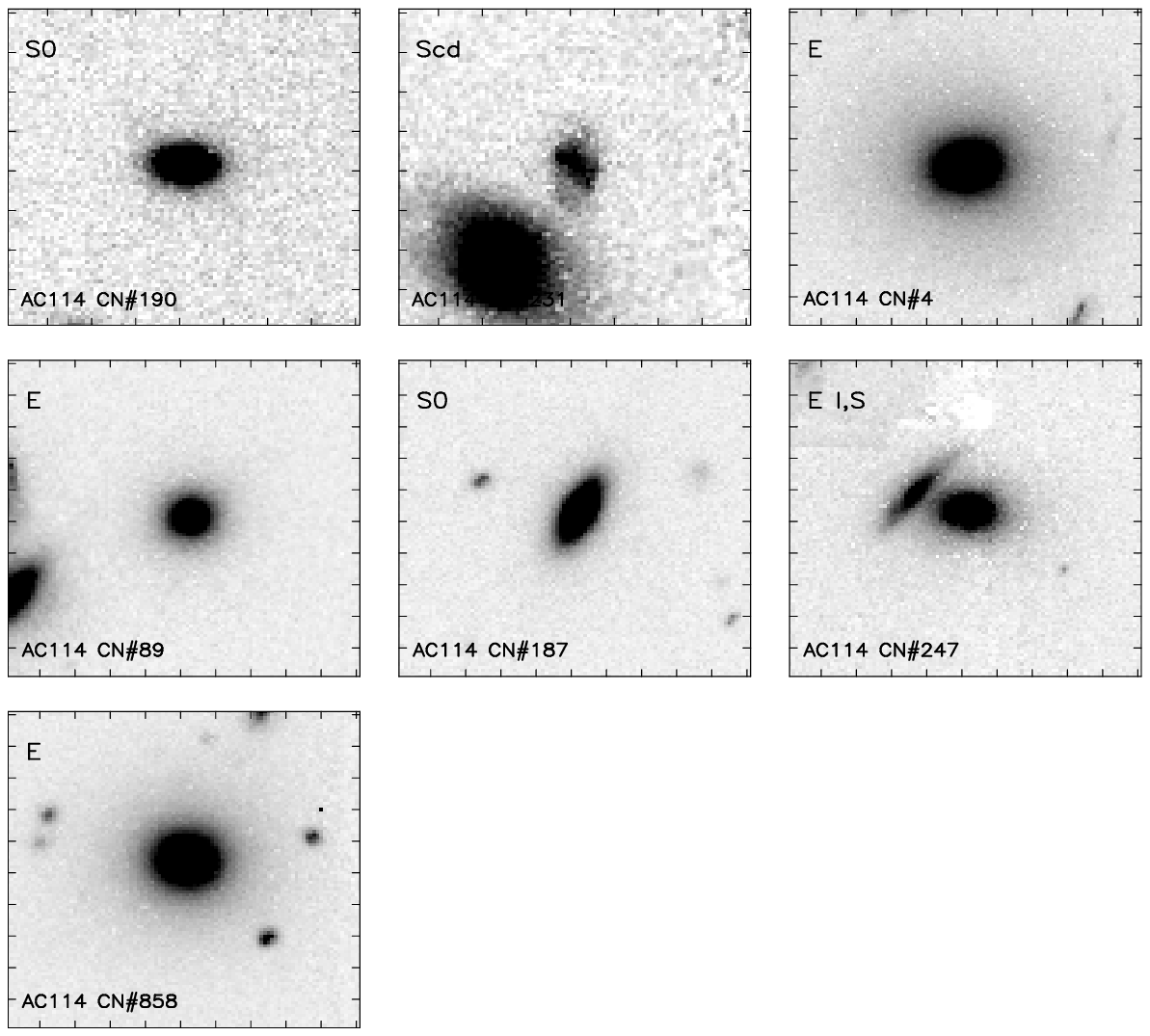,height=240mm,width=204mm}}}
\caption{(c)\,``HDS'' types.}
\end{figure}

\setcounter{figure}{5}

\begin{figure}[p]
\centerline{{\psfig{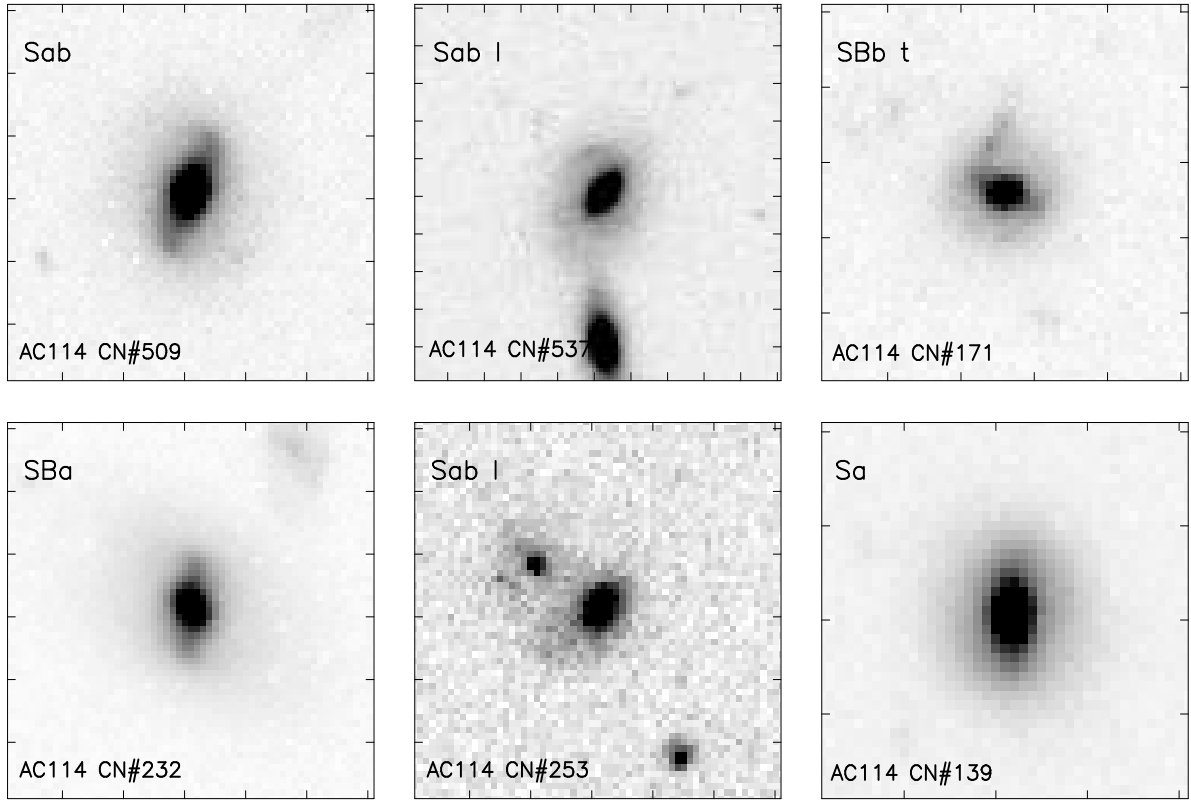}}}
\vspace{-3in}
\caption{(d)\,Spirals with a ``Psv'' classification.}
\end{figure}

\begin{figure}[p]
\centerline{{\psfig{file=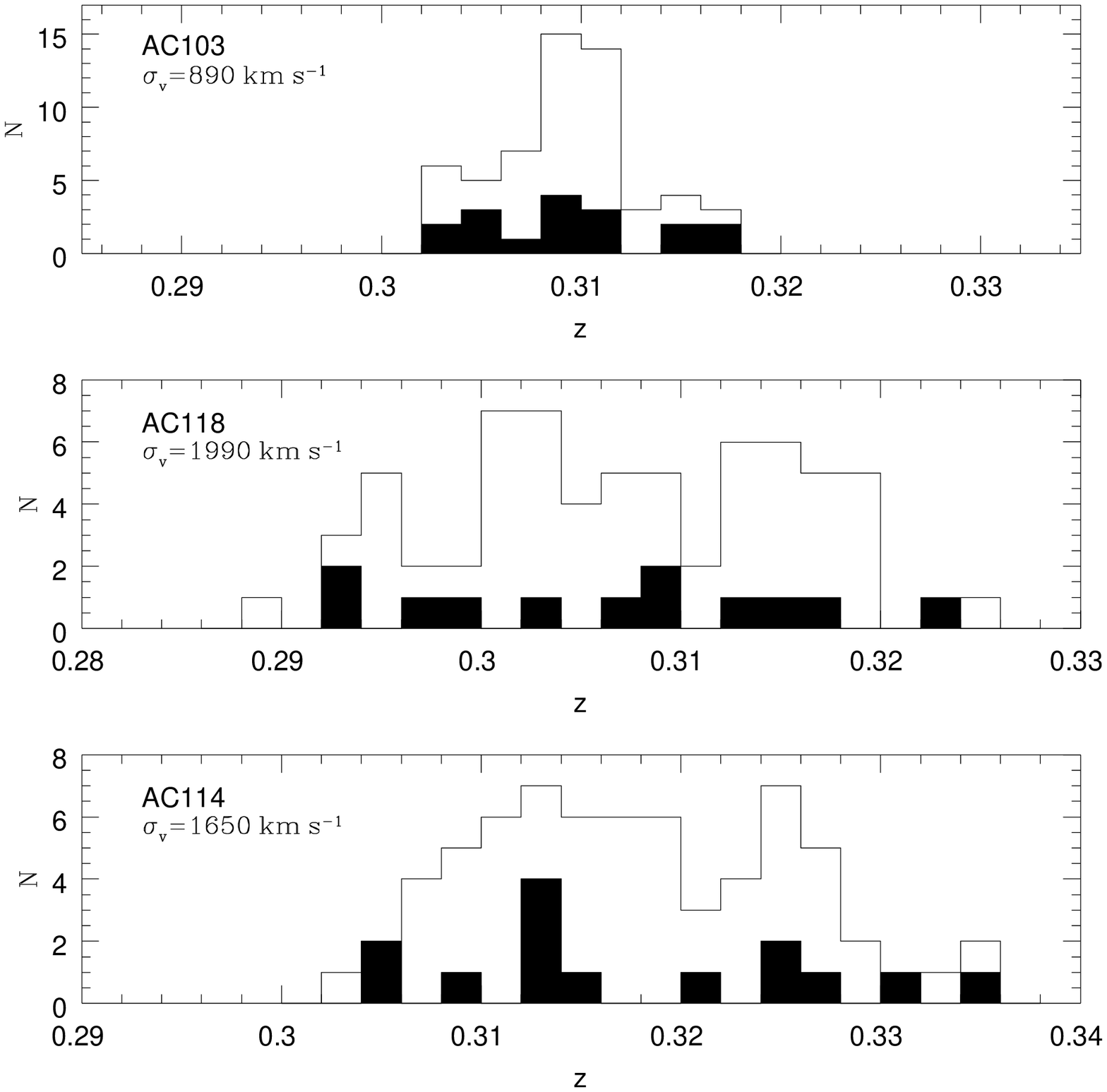,height=240mm,width=204mm}}}
%\vspace{-3in}
\caption{Redshift distributions for our 3 clusters. {\it Open histograms}: all cluster members; {\it solid histograms}: ``blue'' members (ie. those in the ``SB'', ``PSG'' and ``Sp'' spectroscopic classes).}
\end{figure}
               
\end{document}

